\documentclass{article}
\usepackage{PRIMEarxiv}
\usepackage[colorlinks=true,linkcolor=red,urlcolor=blue,citecolor=blue,bookmarks=true]{hyperref}
\usepackage{lipsum}
\usepackage{amsfonts}
\usepackage{subfig}
\usepackage{graphicx}
\usepackage{epstopdf}
\usepackage{amsmath}
\usepackage{amssymb}
\usepackage{cleveref}
\usepackage{amsopn}
\usepackage{setspace}
\usepackage{braket}
\usepackage{color}
\usepackage{algorithmicx}
\usepackage[ruled]{algorithm}
\usepackage{algpseudocode}
\usepackage{algpascal}
\usepackage{footnotebackref}
\usepackage{threeparttable, tablefootnote}
\usepackage[table]{xcolor}
\usepackage{amsthm}
\usepackage{latexsym}
\usepackage[mathscr]{euscript}

\DeclareMathAlphabet\mathbfcal{OMS}{cmsy}{b}{n}

\usepackage[utf8]{inputenc} 
\usepackage[T1]{fontenc}    
\usepackage{url}            
\usepackage{booktabs}        
\usepackage{nicefrac}       
\usepackage{microtype}      
\usepackage{fancyhdr}       
\graphicspath{{media/}}     

\providecommand{\JEL}[2]
{
  \small	
  \textbf{\textit{JEL Codes:}} #2
}

\title{Improving the Efficiency of Payments Systems Using Quantum Computing\thanks{The views expressed in the paper are solely those of the authors and do not necessarily represent the views of the Bank of Canada and other affiliations.  The authors wish to thank Alex Koszegi, Tim Doyle, Peguy Pierre-Louis, and Jonathan Garven from D-Wave Systems for their contributions to the project. We also thank Maryam Haghighi for her inputs and Noorain Noorani for his assistance to the project. We also thank Narayan Bulusu, Rodney Garratt, Victoria Goliber, Isaiah Hull and Hossein Sadeghi for their comments.  $^\ddag$Corresponding authors: deaj@bankofcanada.ca, jplam@uwaterloo.ca} 
}

\author{
  Christopher M\textsuperscript{\underline{c}}Mahon \\
  GoodLabs Studio, University of Waterloo \\
  \And
  Donald McGillivray \\
  GoodLabs Studio \\
  \And
  Ajit Desai$\ddag$ \\
  Bank of Canada \\
  \And
  Francisco Rivadeneyra \\
  Bank of Canada \\
  \And
  Jean-Paul Lam$\ddag$ \\
  GoodLabs Studio, University of Waterloo \\
  \And
  Thomas Lo \\
  GoodLabs Studio \\
    \And
  Danica Marsden \\
  Bank of Canada \\
    \And
  Vladimir Skavysh \\
  Bank of Canada \\
}

\begin{document}
\maketitle

\begin{abstract}
High-value payment systems (HVPSs) are typically liquidity-intensive because the payment requests are indivisible and settled on a gross basis. Finding the right order in which payments should be processed to maximize the liquidity efficiency of these systems is an {\it NP}-hard combinatorial optimization problem, which quantum algorithms may be able to tackle at meaningful scales.
 We develop an algorithm and run it on a hybrid quantum annealing solver to find an ordering of payments that reduces the amount of system liquidity necessary without substantially increasing payment delays. Despite the limitations in size and speed of today’s quantum computers, our algorithm provides quantifiable efficiency improvements when applied to the Canadian HVPS using a 30-day sample of transaction data. 
  By reordering each batch of 70 payments as they enter the queue, we achieve an average of Can\$240 million in daily liquidity savings, with a settlement delay of approximately 90 seconds. For a few days in the sample, the liquidity savings exceed Can\$1 billion.
  This algorithm could be incorporated as a centralized preprocessor into existing HVPSs without entailing a fundamental change to their risk management models.
\end{abstract}

\JEL J{C61, D83, E42, E58}

\section{Introduction}

\noindent High-value payments systems (HVPS), used to settle transactions between large financial institutions, are part of the core financial infrastructure of every country. Central banks that operate or oversee these systems are mandated to ensure their safety and efficiency. Most HVPSs around the world---including Canada's Lynx, the US's Fedwire, and the Eurosystem's TARGET2---are real-time gross settlement (RTGS) systems that settle each payment request on an individual basis, i.e., without netting the offsetting positions of participants. These systems are liquidity-intensive: participants are required to have liquidity before payments are processed.\footnote{To get a sense of their importance, Canada's HVPS processes payment values equivalent to annual GDP every week. In 2021, Canadian financial institutions participating in HVPS allocated Can\$15-20 billion in daily liquidity to process around
forty thousand payments each day.} Liquidity has an opportunity cost, incentivizing participants to wait for incoming payments before making their own. This behaviour reduces the efficiency of a system and can cause delays in payments and even gridlock~\cite{BechGarratt03}. 

Given the systemic importance of HVPSs, improving their efficiency is an important area of research for central banks and academics alike \cite{martin2008liquidity,rivadeneyra2020liquidity, Alexandrova2022Intraday}. 
Improving the liquidity efficiency of an HVPS can be done in two ways. The first is to change the incentives of the participants for the timing of payment submissions, so that payments are more coordinated. The second, once payments have been submitted to the system, is to change the order in which  payments will be settled into a sequence with lower liquidity demand. Inherently, reordering payments requires some delay in payments.

In this paper we take the second approach. We propose a novel algorithm that can reduce the liquidity requirements in HVPSs without significantly increasing the delay of payments. Our algorithm searches over the space of orderings of payment requests. Reordering payments leads to an {\it NP}-hard combinatorial optimization problem that grows to an enormous size with just a few outstanding payments $\left(O(n!)\right)$.\footnote{Our problem has \emph{factorial} time complexity similar to that of the well-known traveling salesman problem, which is an {\it NP}-hard combinatorial optimization problem~\cite{junger1995traveling}.}  Quantum techniques, though not expected to be able to solve such problems in polynomial time, are nonetheless well suited to tackle this problem by providing speedup via heuristics, leading to near-optimal solutions. Following recent technical advances, 
quantum computing, especially when combined with classical computing in a hybrid algorithm, is reaching a stage where it can provide superior solutions to some of these problems compared with classical-only approaches~\cite{NAP25196,Egger20,hull2020quantum}.

Economically, the proposed algorithm can be interpreted as a centralized payments pre-processor. In this setup, participants submit their payment requests to this algorithm before submitting them to the actual system. Subsequently, after the optimization using quantum computing, our algorithm returns to the participants a suggested ordering of payments that reduces the total liquidity required to settle those payments relative to settling them in the original order in which they were submitted. While participants could effectively veto the proposal by not submitting their payments in the suggested order, in practice, if the suggested re-orderings almost always reduce aggregate liquidity demands, participants are unlikely to do so strategically.\footnote{Participants chose their daily liquidity allocation using risk-management models attempting to cover a range of scenarios of liquidity needs. In other words, participants' choices are driven by expected daily liquidity needs \cite{BoE2021chaps}. On an intraday basis, liquidity is rarely managed by choosing specific payments to delay. Likewise, we think it is unlikely that participants would evaluate the algorithm proposal batch by batch.} The proposed setup as pre-processing algorithm provides a simple way to incorporate such algorithms into existing payment systems.

We formulate the reordering of payments in an HVPS as a mixed binary optimization (MBO) and translate it into a quadratic unconstrained binary optimization (QUBO) format for processing on a generic quantum device.\footnote{MBO problems arise in cases with discrete and continuous variables or in the cases with inequality constraints; in our case, we have inequality constraints (see~\autoref{eq:H}). Also, transformation to QUBO is necessary due to the binary nature of quantum bits (or qubit) states and their physical coupling. See~\cite{Lucas2014,Glover2019,hull2020quantum} for further details on QUBO formulation. Note the qubit is the fundamental unit of quantum computing---the quantum version of the classic binary bit.} Due to the size of the problem (pertaining to the use of realistic dataset) compared with the current quantum annealer topology (number and connectivity of qubits), the final problem is posed as a constrained quadratic model (CQM) and optimization performed on D-Wave Systems' infrastructure using their hybrid quantum solver for CQMs~\cite{DWave}. 
The algorithm is run on a sample of 30 days of non-urgent transactions from the Canadian HVPS.\footnote{Note, we use randomly chosen days and settlement data from LVTS---Canada's HVPS before Lynx.}

Our algorithm significantly improved the payments system's efficiency by reducing the required liquidity to process payments with minimal added settlement delay. The input for the optimizer is a given number of queued payments (a ``batch size"). While the set of potential re-orderings grows with the batch size---potentially offering greater improvements in liquidity efficiency---a larger batch size also requires a longer wait to accumulate the payments into the batch and a longer processing time for the algorithm. Among a variety of queue sizes, a batch size of 70 payments strikes a balance between optimizing liquidity savings and minimizing introduced delay in settlements. With that batch size, we find liquidity savings in 26\% of them, providing an average daily savings of C\$240 million. On average, the added settlement delay is 90 seconds, mainly due to the time required to accumulate the batch of 70 payments (the time required to run the algorithm in this case is only 5 seconds). On half of the days in the sample, savings exceed C\$107 million, and on three days, C\$997 million. We also observe that the total liquidity saved for each participant is proportional to the participants' total incoming and outgoing transaction values.

The batch size is constrained by the current size and speed limitations of D-Wave's CQM hybrid solver used in this analysis.\footnote{The CQM hybrid solver can handle a model with 100,000 constraints and 500,000 binary and integer variables~\cite{DWave}.} To test capabilities of different batch sizes, we run the algorithm on two typical days in our sample using batches of 140 payments. We achieve significantly higher savings on those days (C\$326 and C\$94 million above the savings of the 70-payment batch size) with an average delay of 3 minutes per batch (to accumulate the batch of 140 payments). This suggests that liquidity savings could increase substantially as the capabilities of quantum computers scale, and that the more relevant limitation  for this algorithm could be the time required to accumulate larger batches of payments.\footnote{In our case, the time required to accumulate payments grows linearly with batch size (see~\autoref{fig:queSize_waitTime2}).}

We proceed as follows. In Section~\ref{sec:litreview}, we discuss the related literature. We discuss the methodology in Section~\ref{sec:methodology} and  present an overview of RTGS payment systems and the schematic of the proposed algorithm as a centralized payment preprocessor. In the same section, we present the payments queue optimization algorithm and discuss the quantum annealing based hybrid solver used to solve this optimization problem. 
In Section~\ref{sec:data} we provide a brief overview of the sample of the data from Canada's HVPS used to test the quantum algorithm. The results and discussion follow this in Section~\ref{sec:results}, where we investigate the performance of our algorithm. Finally, in Section~\ref{sec:conclusions} we conclude by discussing the opportunities and challenges of the proposed algorithm and the scope for future research. \autoref{app:H} provides further details of the formulation of the objective function.

\section{Literature Review}\label{sec:litreview}

\noindent
The problem of liquidity intensity of RTGS systems is well-known, and alternatives have been explored, in particular liquidity-saving mechanisms (LSM) \cite{martin2008liquidity,diehl2009liquidity,davey2014has,atalay2010quantifying}. The purpose of LSMs is to reduce the liquidity requirements of an HVPS without the credit risks associated with delay net settlement models\footnote{An alternative type of payment system is a deferred net settlement (DNS). In these systems, payments are submitted by participants and accepted by the system but not settled in that moment. Instead, payment exposures are accumulated and settled at given intervals after calculating the net positions of the participants, therefore creating credit risk. See \cite{norman2010liquidity} for an overview of LSMs.}. LSMs temporarily accumulate payment \textit{requests} in a queue until potential offsetting positions are found and settlement occurs. Theoretically, the efficiency improvements of LSMs depend on the liquidity regime \cite{Martin08,martin2008liquidity}, \cite{JurgilasMartin13}, and empirically, new evidence is emerging that questions their effectiveness for improving the efficiency of the HVPS~\cite{Alexandrova2022Intraday}.

In practice, LSMs are designed using heuristics and evaluated using simulation approaches; therefore, they do not search the entire solution space to find the optimal solution \cite{galbiati2010liquidity,rivadeneyra2020liquidity,Rivadeneyra2022Payment}. Compared to the LSM literature, our approach is more direct and transparent: our algorithm attempts to evaluate the entire space of re-orderings looking for liquidity savings.

A newer strand of literature explores an alternative approach, by taking seriously the incentives of the participants arising from the costs they face for liquidity provision and the benefits of settling their payments. \cite{garratt2019application} proposed a mechanism using the Shapley value cost allocation method. In their mechanism, to ensure that the welfare-maximizing netting proposals are always accepted, participants receive take-it-or-leave-it offers to contribute the needed liquidity. The challenges of this approach are that it requires knowing the valuation of the individual benefits of settling each payment, and calculating Shapley values becomes computationally intensive for large sets of payments. 



Quantum computing technology is evolving rapidly, and there is growing interest in its potential for solving certain classes of problems, such as optimization, faster than classical computers \cite{NAP25196}. Using the quantum properties of superposition, entanglement, and tunneling, quantum computers can search large input parameter spaces faster than any classical computer could even theoretically achieve \cite{steane1998quantum,Hidary19}.
Researchers have begun exploring using quantum computing to solve various problems arising in economics---including queuing problems that arise in financial markets~\cite{Egger20,hull2020quantum}.
Closely related to our application, \cite{Braine2021} extend an algorithm for an MBO problem applied to transaction settlement. They optimize small batches of securities trading transactions using gate-based quantum computing. However, their optimization focuses on determining the number of transactions that can be settled while allowing for netting, when the latter does not apply to RTGS systems.




\section{Methodology} \label{sec:methodology}

\noindent
In this section, we first provide an overview of a typical HVPS to understand the incentives of participants and argue how a pre-processing algorithm could be incorporated into the settlement process.  Next, the rationale for choice of computational resources is described.  We then show how such an optimization problem can be generically solved using quantum computing resources, alone or in conjunction with a hybrid solver incorporating classical resources too.  We present the most general QUBO form of the objective function, which could be used by any quantum or classical solver, as well as the input format required for the specific CQM hybrid annealing solver we opted to use due to current physical limitations of quantum computing hardware.  Lastly, we describe the outputs of the CQM solver, and how these outputs are converted to a final solution for the pre-processing of the payments in a given batch.

\subsection{High-Value Payment Systems}



\noindent
In an HVPS, the participating institutions process the payment requests received from their clients. To do so, they apportion collateral to the central bank in exchange for the liquidity necessary to settle those payment requests. The system settles the payments in the order they arrive to the system if they satisfy risk controls. The efficiency of these systems is determined by the amount of liquidity that the participants choose to allocate to the system and by the timing of their payment requests. Participants can delay submitting their payments to await for incoming payments to fund their own payments, reducing the need for their initial liquidity allocation. Such incentives create a trade-off between liquidity and delay \cite{BechGarratt03,castro2020estimating}. Our algorithm seeks to find better solutions to the liquidity management problem by optimizing the order of payments without significantly increasing the settlement delay.

\begin{figure}[!h]
  \centering
  \includegraphics[width=.99\textwidth]{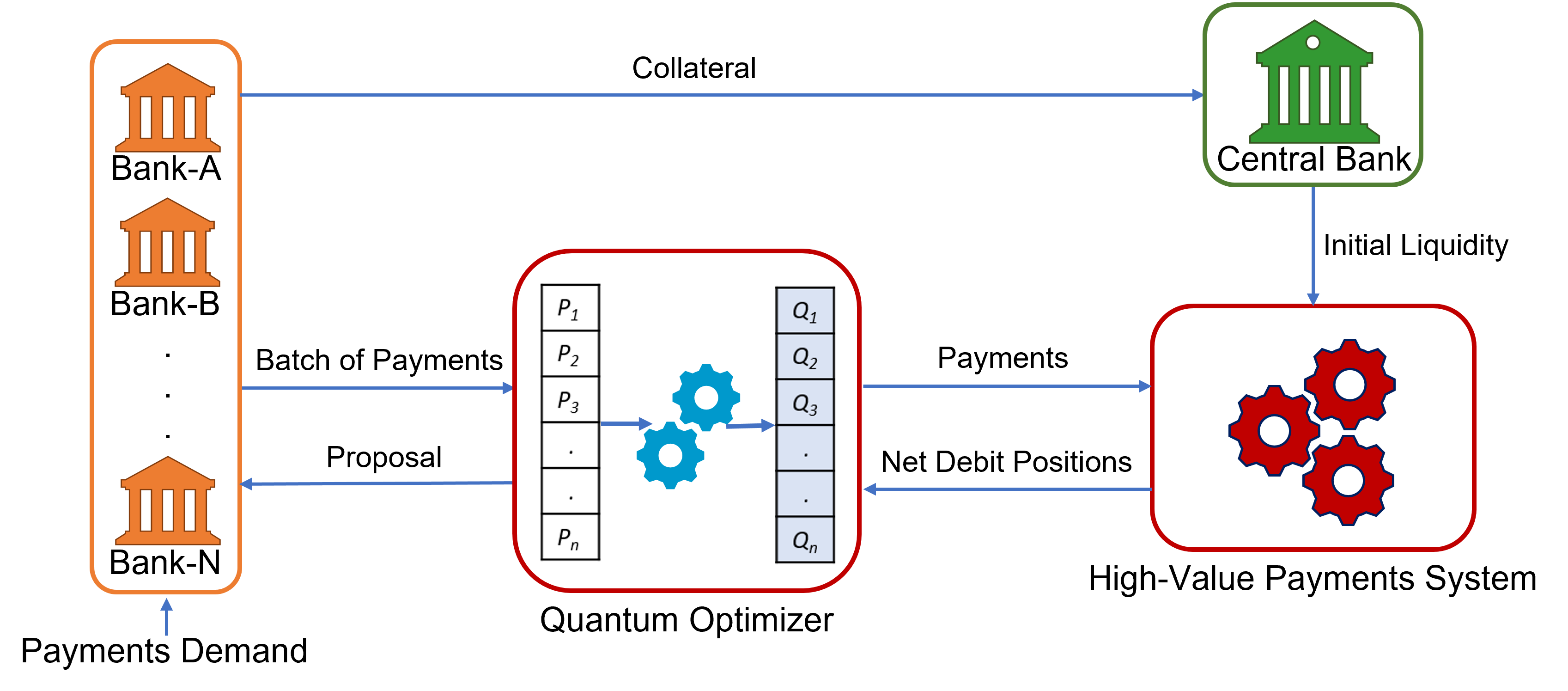}
    \caption{Schematic of the quantum optimizer as a pre-processor to the payment system. There are $N$ participating banks; they receive multiple payment requests from their clients throughout the day. Banks submit payments to the central queue in the order $\{P_1, P_2, \dots, P_n\}$. The quantum optimizer then processes those payments and proposes a new order $\{Q_1, Q_2, \dots, Q_n\}$. If each bank accepts the proposal, those payments will be submitted to the HVPS. Each participant's initial and intraday liquidity positions (updated after collateral apportionment at the start of the day or after payment settlement throughout the day) are provided to the quantum optimizer by the payment system. }
  \label{fig:quantum_preprocessor}
\end{figure}

~\autoref{fig:quantum_preprocessor} is a stylized schematic of a wholesale payment system. The figure also shows a generic quantum optimizer as a central pre-processing mechanism between participants and the payment system. A key challenge for new algorithms seeking to improve the efficiency of existing payments systems is how to incorporate them without incurring a fundamental change to the system. As a pre-processor, quantum optimization algorithms could be used to propose orderings that tangibly benefit the participants, incentivizing them to submit the payments to the system in the order suggested by the algorithm.

Therefore, in our approach, each participant submits payment requests to a centralized queue denoted in the ordered set $\{P_1, P_2, \dots, P_n\}$. Each element in this set contains information about the value of the payment, the two participants involved (payer and payee), and the time the request was submitted. Once there are enough payments in the queue (a batch), the quantum optimizer processes those payments and provides a proposal for an optimized queue in the order $\{Q_1, Q_2, \dots, Q_n\}$ that minimizes liquidity requirements and contains the new submission timing. If all banks accept the proposal, those payments are submitted to the system in the new order proposed by the quantum optimizer. As payments are settled throughout the day, each participant's liquidity position (balance) changes; this information---necessary for the optimization process---is fed to the quantum optimizer by the payment system.

Our approach is to have a fixed size $n$ of the batch of payments that will be optimized by the quantum optimizer. Once the number of payments received reaches that size, the optimization is triggered. Since the flow of payment requests varies though the day, the optimization occurs with varied frequency (see ~\autoref{fig:queSize_waitTime2}). 
To evaluate the algorithm we report delay statistics that consider the time needed to collect enough payments (wait time) for the batch and the time required by the combined compute resources---a hybrid of quantum and classical---to evaluate the optimized queue (processing time). Given the wait time, we choose to test the algorithm on non-urgent payments.\footnote{Our approach with a fixed batch size is potentially not suitable for urgent payments, because it requires some wait time until a batch of payments of a given size has accumulated. This problem can be alleviated to a certain extent using flexible batch sizes. In the Canadian HVPS, however, urgent payments typically represent only about 1.5\% of the total volume of payments, and they are typically very large in value and usually come at specific times of the day.}

\subsection{Quantum Formulation of Queue Optimization Algorithm} \label{sec:qubo1}

\noindent
The goal of the optimization is to minimize the aggregate liquidity needed in the system to settle a given set of payments, ordered with index $i=1\ldots n$. To measure the liquidity used in the system, we start by defining $N(i)$ as a participant's net position before a payment $i$ is settled. A participant's net position is their liquidity balance accounting for all incoming and outgoing payments up to that time in the day. 

Next, we define $mNDP$ as the maximum net debit position experienced by a participant at any time thus far in the day. If a participant begins the day with $N=\$0$, then the end-of-day $mNDP$ after all payments have settled measures the minimum liquidity a participant would have required to process its payments without incurring a negative position. We can therefore think of $mNDP$  as equivalent to a participant's minimum liquidity requirements. Reducing the sum of $mNDP$  across all participants, holding payment values and settlement delay fixed, represents an improvement in the liquidity efficiency of the system. Further, let $b$ represent a participant's liquidity allocation.

The quantum optimizer pre-processor rearranges the queue of payment instructions with an alternative index $t=1\ldots n$ that meets the goal of minimizing total liquidity needs, subject to the constraints that no participant gets into a negative liquidity position at any point in time and that no payment remains in the queue.

We can construct a Hamiltonian in QUBO format, which aligns the minimum energy of a quantum annealing system to the optimal queue order.  While some problems have a mapping already known in the literature, for example \cite{Lucas2014}, our problem did not.  Thus we derived a formulation in~\autoref{app:H} for a QUBO which is the most general form of a problem which can be easily run on any type of current quantum computer or even classical solvers\footnote{We use $H$ to represent the objective to be minimized over, as in quantum annealing it is the energy being minimized, analogous to applying the Hamiltonian operator to the state space.  Note that this Hamiltonian is slightly different from the one encountered in economic applications of optimal control theory, which often contains a discount factor and hence needs transformation, and the new operator is commonly referred to as the \emph{current-value} Hamiltonian~\cite{chiang1992hamiltonianEcon,cass2014hamiltonian}.}
\begin{align}
H=& \sum_\alpha b_\alpha + \lambda_1\sum_{\alpha,t} \left(b_\alpha + N_\alpha(0) + {mNDP}_\alpha + \sum_i\sum_{\tau\leq t} f(\alpha,i)\, x_{i,\tau} - s_\alpha(t) \right)^2 \nonumber \\
& + \lambda_2\sum_i \left( 1-\sum_t x_{i,t}\right)^2 + \lambda_2\sum_t\left(1-\sum_i x_{i,t} \right)^2,
\label{eq:H1}
\end{align}

\noindent
where:

\begin{tabular}{r p{0.75\textwidth}}
     $b_\alpha$  & is the participant $\alpha$'s liquidity allocation.  When minimized, it is the amount of additional liquidity that $\alpha$ would need to have to permit settlement of the payments in that batch (i.e., to avoid gridlock). Equivalently, it is the increase of $mNDP_\alpha$ after this batch is settled  \\
     $N_\alpha(0)$  &  is participant $\alpha$'s net position immediately before this batch is settled  \\
     $mNDP_\alpha$  &  is participant $\alpha$'s maximum net debit position incurred during previous batches earlier in the day  \\
     $x_{i,t}$  &  is a binary decision variable that indicates whether payment $i$ is settled ($x_{i,t}=1$) or not ($x_{i,t}=0$) at position $t$ in the final queue  \\
     $s_\alpha(t)$  &  is a non-negative real number acting as a slack variable that permits the enforcement of an inequality  \\
     $f(\alpha,i)$  &  is a function that returns $v$ (the value or amount of payment $i$) if $\alpha$ is the payee; $-v$ if $\alpha$ is the payer; and 0 otherwise  \\
     $\lambda_1$, $\lambda_2$  &  are Lagrange multipliers that enforce constraints as quadratic penalty terms.  \\
\end{tabular}

\noindent
Minimizing $H$ over the set of variables $\{b_\alpha,x_{i,t},s_\alpha(t)\}$ $\forall\alpha,i,t$ finds the optimal ordering $\{x_{i,t}\}$ that requires the least increase to the aggregate $mNDP$. Note that the optimal ordering may be many times degenerate.


\subsection{Quantum Annealing and Hybrid Solvers}
\label{sec:cqm1}
\noindent
To solve an optimization problem, several tested commercial and open-source optimization software packages represent the state of the art.  They implement pre-solvers and heuristics to simplify the problem and find a solution (SCIP, CPLEX, and Gurobi are some of the most popular; see  \cite{BestuzhevaEtal2021ZR, Vigerske2017, Rimmi2017}). In addition, they greatly simplify many technical issues in solving these optimizations, especially the non-linear restrictions, and allow for greater focus on their modelling.  However, as stated earlier, for problems that scale with the number of variables as ours does, classical computing resources begin to fall short; hence the desire to explore quantum methods. 
 
A variety of different technologies are being explored as the best means to build quantum hardware, but they are primarily classified into universal gate-based quantum computers, which can process logic and run general algorithms, and quantum annealers, which are used exclusively to solve optimization problems. While both technologies can theoretically solve the optimization problems explored here, currently available quantum annealers can solve much larger optimization problems than gate-based quantum computers can, and their hybrid solver counterparts larger problems still~\cite{DWave}.
 
D-Wave Systems' latest quantum annealer (also known as a quantum processing unit, or QPU), accessible via the cloud, has more than 5,000 qubits with 15-way qubit-to-qubit connectivity (the ``pegasus" topology).  In addition to their quantum annealer, D-Wave Systems offers access to 3 hybrid solvers which make use of both classical and quantum compute resources: the constrained quadratic model (CQM), binary quadratic model (BQM), and discrete quadratic model (DQM). Each of these solvers allows for different levels of flexibility in the construction of the models, numbers and types of variables, and numbers of constraints.  

In the case of the variable matrix mapping $i$ to $t$, the number of variables to be considered here grows as the square of the batch size, quickly overwhelming the size of the QPU.  Thus, for our problem, as the number of payments to be reordered increases, even 5,000 qubits with 15-way connectivity (which equates to around 180 fully connected qubits or variables) are not enough to embed the problem in order to perform the anneal.  Since the CQM hybrid solver can handle more constraints (100,000) and binary and integer variables (500,000) than any other available quantum hardware, we opted to use it.\footnote{The main downside to using CQM hybrid solvers is that the details of their inner workings are not known. On D-Wave Systems' QPU, parameters such as annealing time, number of reads, and annealing path can be controlled.  For the hybrid solvers, these parameters are tuned automatically, and the only input parameter (optional) is a time limit for the calculation. Aside from the returned solutions, only the amount of QPU time used per calculation can be retrieved.}

The CQM solver takes objectives and constraints directly as inputs; thus, the problem was submitted specifically as:
\begin{align}
\text{Objective:} \qquad &  \text{min} \sum_\alpha b_\alpha \\
\text{Subject to:} \qquad &   b_\alpha + N_\alpha(0) + {mNDP}_\alpha + \sum_i\sum_{\tau\leq t} f(\alpha,i)\, x_{i,\tau} \geq 0, \forall \alpha   \label{eq:constraint3} \\
 &   \sum_i x_{i,t} = \lambda, \forall t\\ 
 &   \sum_t x_{i,t} = \lambda, \forall i
\end{align}

\noindent negating the need for slack variables and one Lagrange parameter on the user's end.  The explanation of variables from Section \ref{sec:qubo1} otherwise remains the same, including that $x_{i,t}$ are binary decision variables.

The economic intuition of the problem is neatly summarized by the main constraint of the problem (\autoref{eq:constraint3}). There, an increase in the first term indicates a deterioration in the solution to the problem by requiring additional liquidity in the system above the $mNDP$ observed up to the processing of that batch. The last term in that constraint is the value of settled payments, which is the product of the variable that indicates if a payment has settled or not ($x_{i,t}$) and the function of the value of the payment ($f$). Although this term changes discretely, it is useful to think in marginal terms. A marginal increase in this term indicates that more payment value is settled, which, if the constraint binds, can only be satisfied by increase in liquidity, worsening the solution. Therefore, by searching over different sets $\{x_{i,t}\}$, the optimizer explores the trade-off between settling more payment value and increasing the liquidity necessary to do so.  

Note that, using the CQM solver, each batch is processed as if it were happening in real time, and only analyzed to yield aggregate results afterwards.  That is, we do not look ahead to use knowledge of what would come later in the day to organize each earlier batch.  Thus, our method, while true to what could be achieved in real-world scenarios, optimizing batch by batch, doesn't always result in achieving the minimum liquidity requirements over the course of a full day. To achieve these would require, among other things, predicting which payments may appear in subsequent batches.  As we will show, on average, institutions are nonetheless still positively affected, proportional to the value of the transactions they conduct.

\subsection{From CQM Solver Results to the Optimal Queue Order}
\label{sec:postproc}

\noindent
Finding a solution amounts to finding the matrix that represents the reordering of the payments from $i$ to $t$, i.e., $\{x_{i,t}\}$. For example, if there are seven payments in a queue, the CQM solver might return:
\begin{equation}
  \{x_{i,t}\} = \begin{pmatrix} 
    0 &  1 &  0 &  0 &  0 & 0 & 0 \\
    0 &  0 &  0 &  0 &  1 & 0 & 0 \\
    0 &  0 &  0 &  0 &  0 & 1 & 0 \\
    1 &  0 &  0 &  0 &  0 & 0 & 0 \\
    0 &  0 &  0 &  1 &  0 & 0 & 0 \\
    0 &  0 &  0 &  0 &  0 & 0 & 1 \\
    0 &  0 &  1 &  0 &  0 & 0 & 0 
    \end{pmatrix}\label{eq:outputMat}
\end{equation}

\

\noindent
which would mean that the fourth payment in the original queue should be processed first, then the payment at the top of the queue, and so on.  To find these solutions, our algorithm takes the payments in the batch, generates the objective and constraints imposed by each payment in that batch, and runs several times, returning a reordering solution $\{x_{i,t}\}$ each time.  

The reason for multiple runs is the nature of quantum computation.  Each run of the annealer potentially ends in a different state based on the length of the anneal (analogous to ``cooling time" in a physical anneal) and the underlying probability distribution of the states (the lower the energy of the final state, the higher its  probability of occurring)\footnote{The problem is run on the quantum annealers by first starting from the state where all qubits are in a superposition of the basis states $\ket{0}$ and $\ket{1}$. This corresponds to the lowest energy state of the initial ``tunneling" or ``mixing" Hamiltonian. This initial Hamiltonian is then evolved to a Hamiltonian encoding the problem that is being solved (in this case given by  \autoref{eq:H1}, enacted through coupling parameters and biases on the qubits. If this change is slow enough, the adiabatic theorem guarantees that the system will end up in the lowest energy state of the final Hamiltonian, which corresponds to an optimal solution to the problem, the number of which corresponds to the degeneracy of the ground state.}.  In practice, time constraints mean that we may reach the near-optimal, but not best, solution in a given run.  However, more solutions provide more confidence based on the statistics of the solution set that one of the optimal reorderings was found (there may be multiple reorderings which are equally good).

Our first step in processing the set of returned reordering solutions is to check if any violated the constraints in the Hamiltonian, i.e., if it is an infeasible solution.\footnote{For instance, if there are more than one non-zero elements in each row or column of the output matrix in~\autoref{eq:outputMat} then that is an infeasible solution.}  This information is provided by the CQM solver.  If so, those infeasible solutions are discarded. The remaining feasible solutions are then searched classically for the ones that provided the most savings, yielding our final result and the payments processing order for that batch. 

The optimality of the quantum solution is challenging to quantify for larger problems, since optimal solutions become too computationally expensive to verify (the reason we turn to quantum computation in the first place). However, an approximate measure of optimality can be inferred from the number of feasible candidate solutions returned from the CQM solver. The CQM solver's default run time for batches of 70 payments, for example, provides about 50-70 solutions.  However, as the problem gets larger with queue size, fewer feasible solutions come back from CQM. This indicates that the quality of the solutions goes down. Thus, if the CQM solver returns a histogram with many (few) feasible results, it should correlate with a better (worse) solution.  Indeed, we observe this correlation (see \autoref{fig:Batch_size_vs_FIFO_r700}).

\section{Data}
\label{sec:data}

To test the performance of the proposed quantum algorithm, we randomly sample 30 days from Canada's LVTS settlement data between Jan 2015 to Dec 2017. From those 30 days, we use only non-urgent payment requests made between 8 am and 6 pm.\footnote{In the LVTS, non-urgent payment requests are submitted to Tranche 2, and urgent payment requests are submitted to Tranche 1. The urgent payments could have a high cost of delay; therefore, we exclude them from the sample. The payment requests submitted after 6 pm and before 8 am are special transactions; therefore, they are also excluded from the chosen sample (see \cite{arjani2006primer}).} In our chosen sample, after excluding non-urgent payments, we have approximately 23,000 transactions per business day. On some busy days, however, the LVTS processed a higher volume of transactions.\footnote{On average, the LVTS processes approximately 37,000 transactions per business day. However, on some busy days, such as the end of the month, quarter, or the day after a national holiday, the LVTS processes about 45,000 to 50,000 transactions.} Therefore, to test the benefits of a quantum optimizer on such days, we ensure that ten high-volume days are included in our 30-day sample.\footnote{Before randomly sampling 30 days, we exclude a few days from our sample on which volume is minimal, for instance, on Canadian provincial or US national holidays.}

\begin{figure}[!h]
  \centering
	\includegraphics[width=0.49\textwidth]{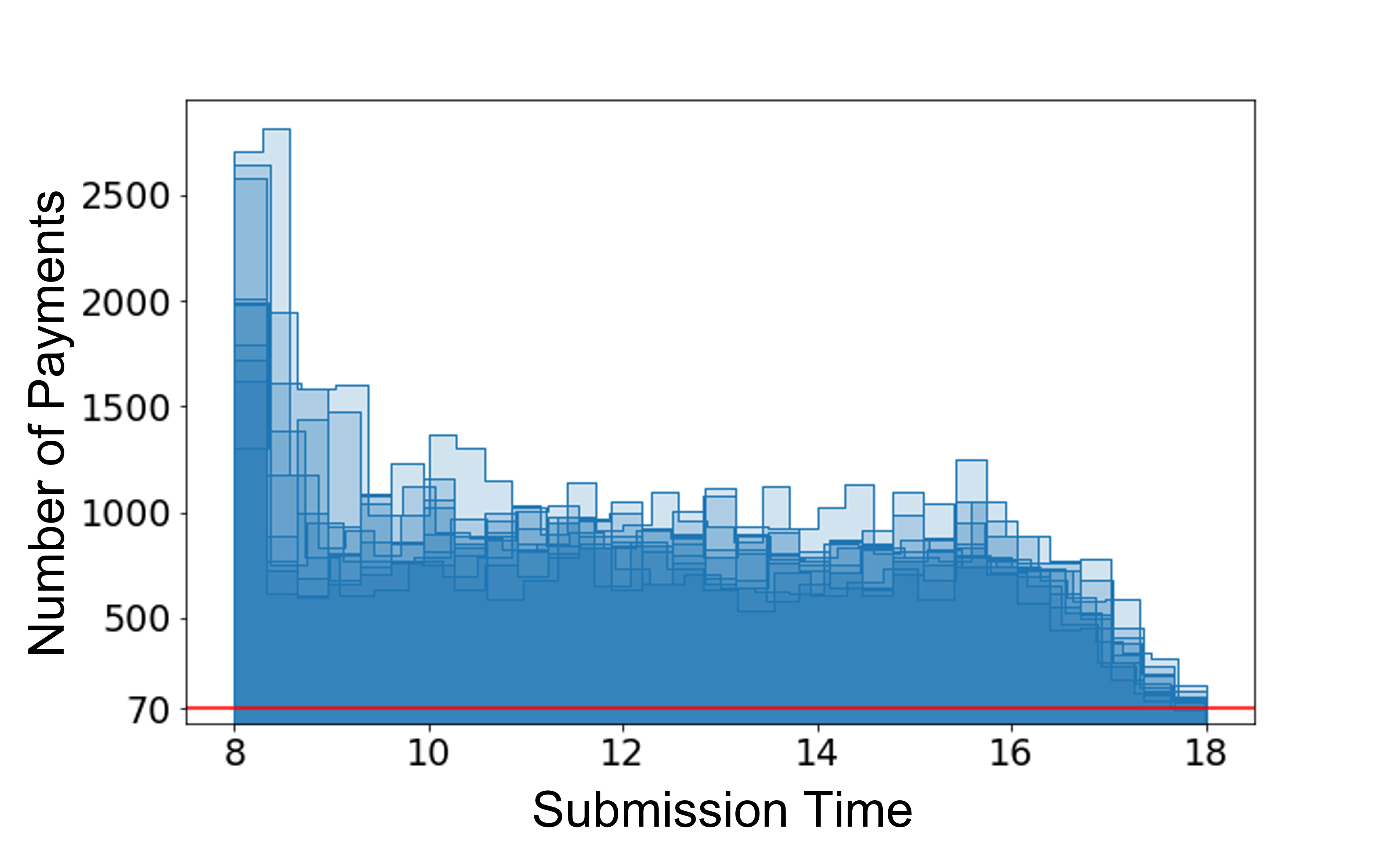}
	\includegraphics[width=0.49\textwidth]{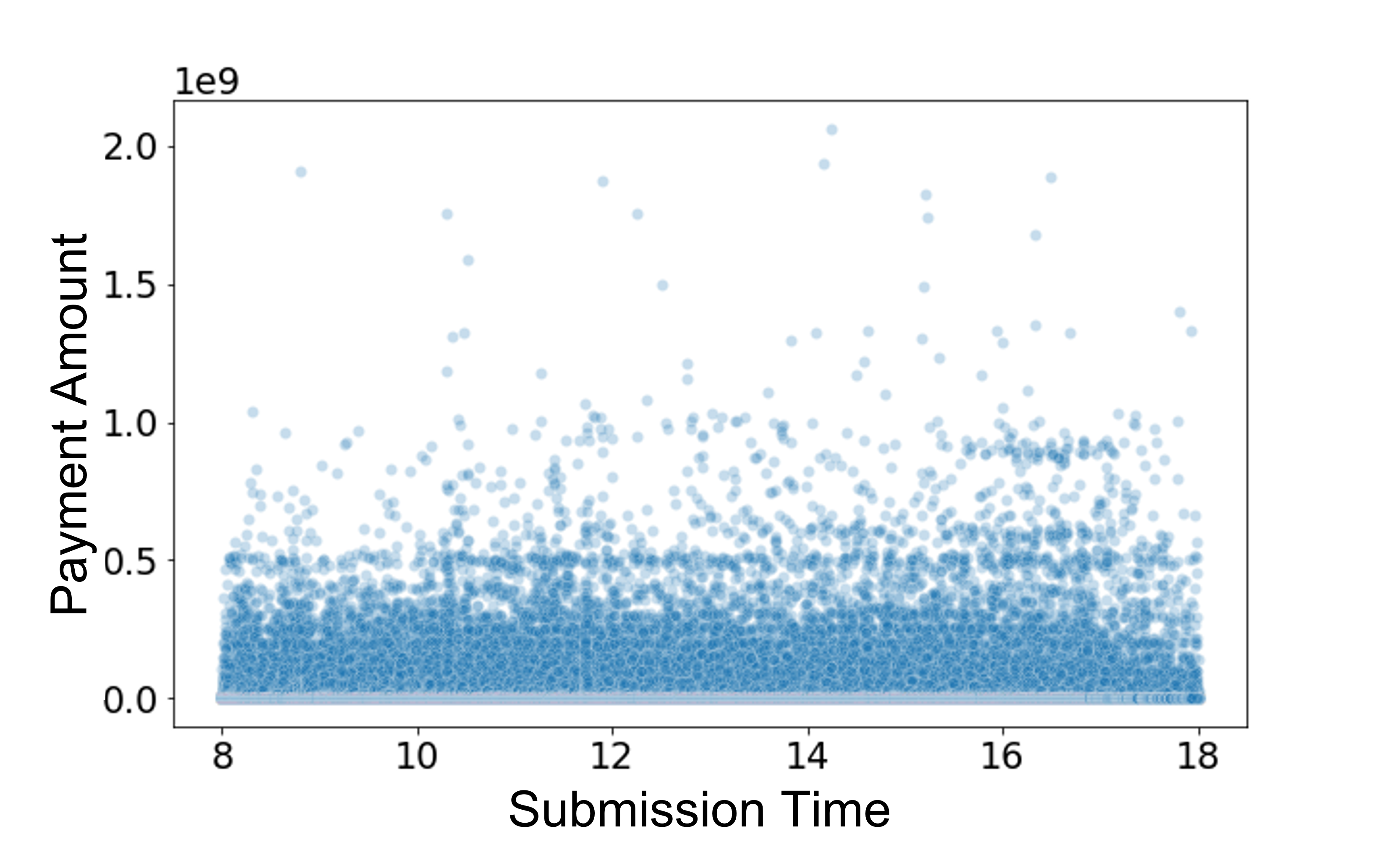}
    \caption{(Left) the number of payments and (right) payment amount submitted to the system over time for each day in our sample.}
  \label{fig:queSize_waitTime1}
\end{figure}

In our sample, the frequency at which payment requests are submitted varies with the time of day, as shown in~\autoref{fig:queSize_waitTime1} (left). We see a much higher volume of payments settled between 8 to 9 am. The number of payment requests remains steady from about 9 am until about 4 pm, before they taper down towards the end of the day. On the other hand, the payment amount submitted to the system varies almost uniformly throughout the day. However, the share of the number of payments with high value is slightly higher towards the end of the day in our sample, as shown in~\autoref{fig:queSize_waitTime1} (right).  

\section{Results and Discussion}  \label{sec:results}
This section presents the results of various simulation exercises performed on the 30 days sampled from Canada's LVTS settlement data using the D-Wave system's CQM hybrid solver. First, we present the results concerning the computational time to accumulate and solve queue optimization problems for different batch sizes. Next, we examine the solution quality of the quantum optimizer. This is followed by the performance of the quantum algorithm on the entire sample. Finally, we investigate the performance of the quantum algorithm for a chosen day in our sample. 

\subsection{Computation Time}
\label{sec:comptime}
\begin{figure}[!h]
  \centering
    \includegraphics[width=0.72\textwidth]{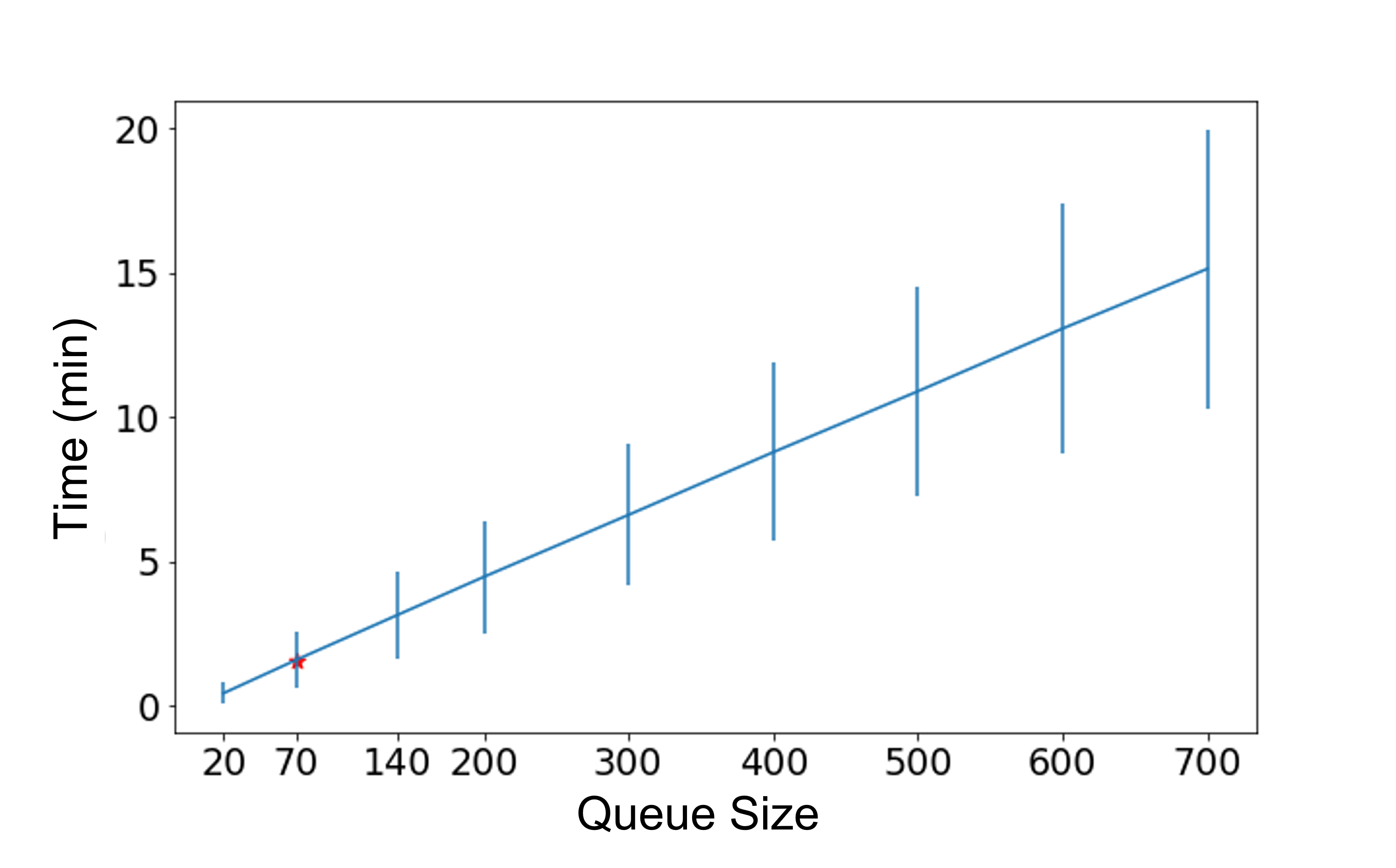}
    \caption{Average waiting time to fill a queue for a given queue size. The orange star highlights the queue size of 70 payments.}
  \label{fig:queSize_waitTime2}
\end{figure}

\begin{figure}[!h]
\centering
\includegraphics[width=0.65\textwidth]{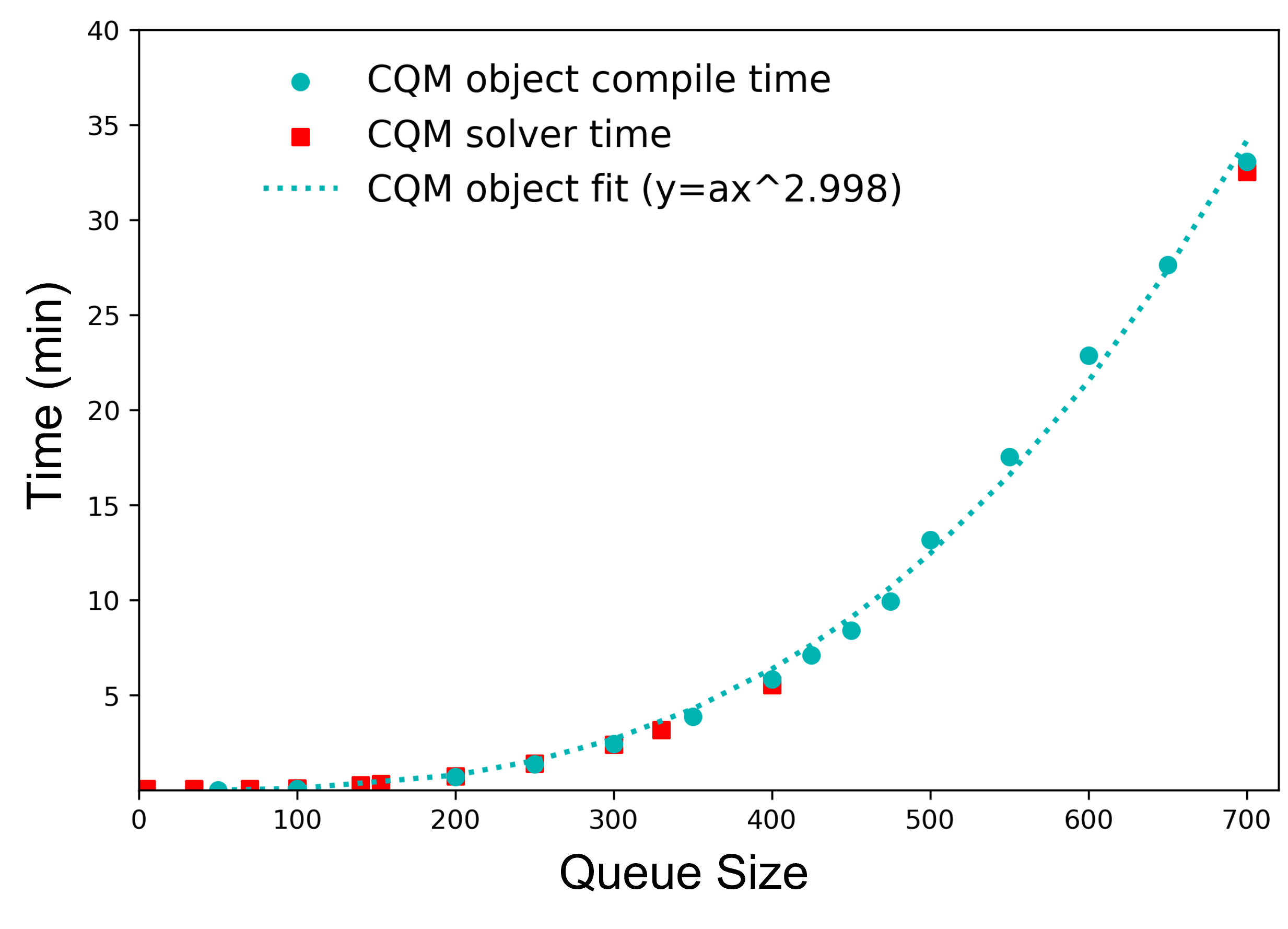}
\caption{Time needed to compile the CQM object (blue) and then find the solution using D-Wave Systems' hybrid CQM solver (red), along with a best fit.  Note that the same day's payments are used to compute the time, but truncated to the desired queue size.}
\label{fig:size_time}
\end{figure}

\noindent
The reordering of a queue using the quantum optimizer consists of the following steps: 
\begin{enumerate}
    \item accumulating the batch of payments in the order they arrive
    \item building the CQM object formulated in Section~\ref{sec:cqm1} 
    \item sending the object to D-Wave Systems' hybrid CQM solver to perform optimization 
    \item post-processing the solver's output to extract the solution. 
\end{enumerate}
For step 1, the wait time to collect a batch of payments grows linearly with the batch size (\autoref{fig:queSize_waitTime2}). However, for steps 2 and 3, the processing time and memory required by the CQM object grow cubically with batch size. \autoref{fig:size_time} shows the time to compile and solve the CQM object versus queue size ranging between 20 and 700 payments. The cubic growth is due to the scaling of the number of biases, which grows like $O(n^3)$ in \autoref{eq:H1}. Finally, in step 4 the CQM solver's output is processed to get the best solution, i.e., the order in which to submit the payments.

For real-world applications of such algorithms, solving the quantum reordering should not take longer than collecting the batch of transactions in order to avoid added delays. For a queue size ranging between 20 and 700 payments, the average wait time ranges from 1 to 15 minutes (\autoref{fig:queSize_waitTime2}). However, this wait time varies with the time of day because the frequency at which payment requests are submitted changes throughout the day (\autoref{fig:queSize_waitTime1}, left). 
For a batch of 70 payments, the average wait time is about 90 seconds, and average compiling and processing time is 5 seconds each.  
The cubic growth in time shown in \autoref{fig:size_time} limits the maximum queue size to around 140 transactions, which takes on average about 115 seconds to compile and solve the problem. However, there is room for potential improvement by calculating the CQM on more powerful computers with multiprocessing 
and reducing network latency by calculating the problem on servers physically closer to the solvers. This could be explored in the future. 


\subsubsection{Classical Comparison}

To compare our optimization with a classical solver, we formulate our algorithm to run on SCIP, a non-commercial solver for mixed-integer nonlinear programming \cite{BestuzhevaEtal2021ZR, Vigerske2017}. It utilizes pre-solving and heuristics to divide the problem into subproblems and tighten domain variables to solve recursively, similar to CPLEX \cite{Rimmi2017}. Unfortunately, depending on the set of transactions submitted, the structure of the network graph, and the mNDP's lower bound, it generally takes more than 2 hours to solve and occasionally fails after more than 24 hours for a batch size of 70, run on a \textit{R5.4xlarge} AWS instance (16 vCPUs at 3.1GHz, 128GB RAM).  This suggests that our problem is better suited to being run on DWave's hybrid quantum and classical solver.

\subsection{Solution Quality} 
\label{sec:solnqual}

\begin{figure}[!h]
    \centering
        \includegraphics[width=0.7\textwidth]{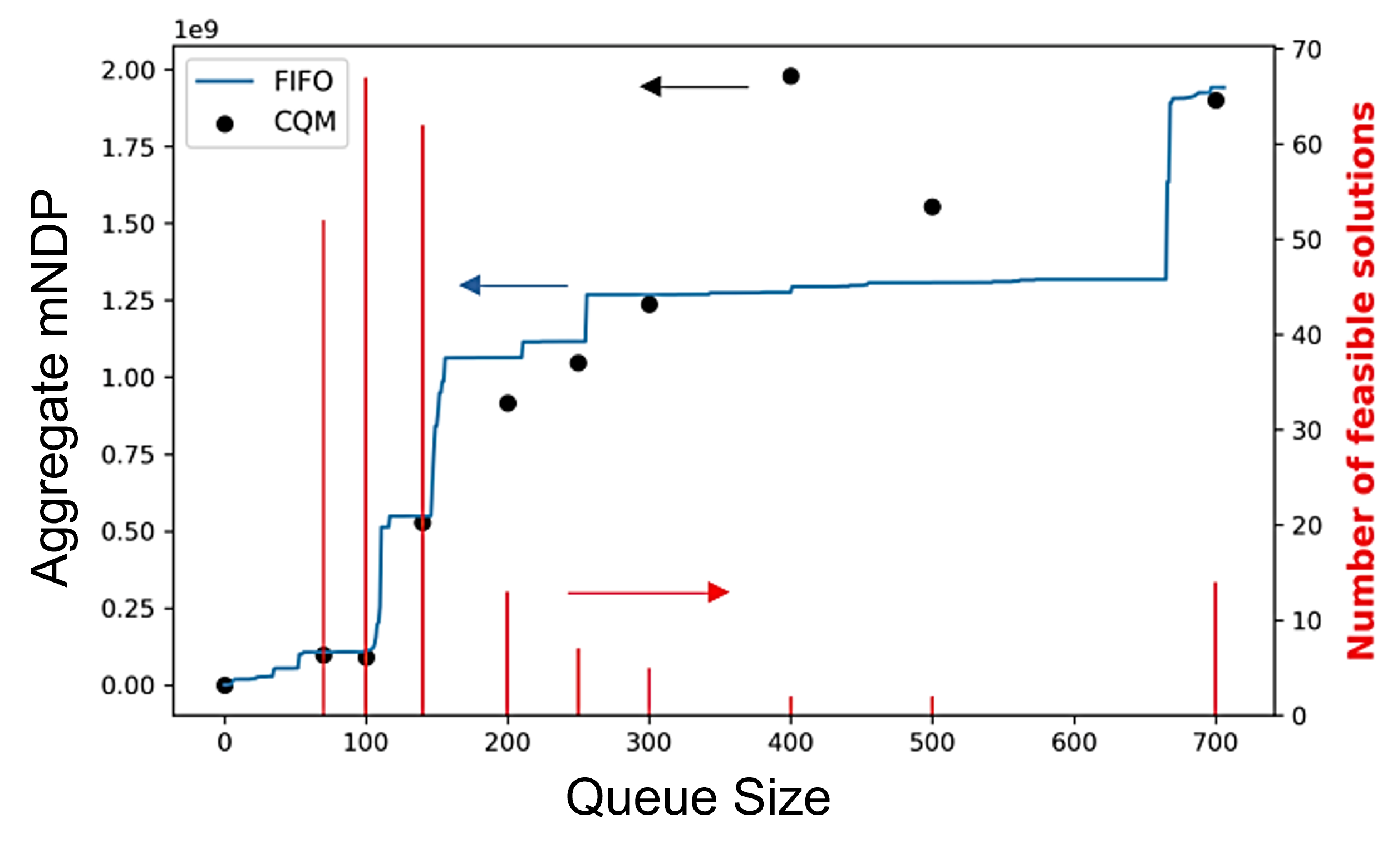}
        \caption{The aggregate maximum net debit position (mNDP; left) and number of feasible solutions (right) with respect to batch size. The mNDP of FIFO is included for reference (blue line). Note that the hybrid CQM solver computation time limit for batches of size 300, 400 and 500 was set to 5, 11 and 26 minutes, respectively (2 times the CQM default setting). For the batch size of 700, a time limit of 3 hours was used (6 times the default).}
    \label{fig:Batch_size_vs_FIFO_r700}
\end{figure}

Using a representative set of 700 payments, \autoref{fig:Batch_size_vs_FIFO_r700} shows the aggregate mNDP versus batch size for single queues containing the first $n$ payments, with $n$ ranging from 70 to 700, (black dots).  The comparison to a FIFO queue arrangement is also shown (blue line), as well as the number of feasible solutions returned by the CQM solver (red bars). Up to and including a batch size of 300, the solution from the CQM solver is better than FIFO; however, the results become worse than FIFO as batch size continues to increase.  This decline in quality of results corresponds to a drastic decrease in the number of feasible results returned.  For batch sizes above 250, the default time limit parameter for the calculation (chosen algorithmically by the CQM solver based on multiple properties of the CQM input) was not enough to obtain any feasible results and had to be overwritten. 

The data point corresponding to a batch size of 700 used 3 hours of hybrid quantum computation time---set manually---to achieve a solution that is superior to FIFO. These results suggest that we have found the maximum problem size ($n\lesssim200$) that current hybrid classical-quantum annealing technology can solve in a reasonable amount of time. However, it also implies that as the technology continues to develop and scale, future hybrid solvers or quantum annealers with more entangled qubits, which can search larger state spaces, will be able to handle larger queues.

\subsection{Performance of Quantum Algorithm on the Entire Sample}

As mentioned earlier, we optimize the settlement order by dividing each day into batches and solving each batch on the CQM solver. These batches are solved using information (each participant's mNDP and net position) from the previous batch. The summary of the results for the entire sample is presented in~\autoref{table:b70_stats_main} and~\ref{table:b70_stats} for a batch size of $n=70$.
Note that these tables represent two different scenarios. For~\autoref{table:b70_stats_main}, we run a horse race between the FIFO and CQM. They both start with the same initial conditions (for instance, the same mNDP at 8 am). Consequently, every batch throughout the day is the same for both FIFO and CQM. At the end of the day (6 pm), we report the results (mNDP). However, for~\autoref{table:b70_stats}, we run the batch-by-batch horse race between FIFO and CQM. In both FIFO and CQM, we start with the same initial conditions for every batch, check which gives a better result, and report those results.


\begin{table}
\centering
\caption{Summary statistics of the sample data showing the total value of payments settled in Tranche 2 between 8 am and 6 pm. The days are sorted in descending order of total number of batches, where each batch contains $n=70$ payments. A batch is considered optimizable if settlement by the original order in which they came, (i.e., FIFO), increases the aggregate mNDP. The improved batches column shows the number in which the quantum optimizer using the CQM solver was able to reduce the aggregate mNDP. The last column shows the total liquidity saved by the end of the day using the CQM solver.}

    \begin{tabular}{| c |  >{\raggedleft\arraybackslash}m{0.15\textwidth} | m{0.1\textwidth} |>{\raggedleft}m{0.1\textwidth} | >{\raggedleft\arraybackslash}m{0.17\textwidth} |}
    \hline
         Date  &  \centering Value Settled (\$Bn)  &  \centering Total Batches  & \centering Improved Batches  &  \centering End-of-Day Savings (\$Mn)\arraybackslash
         \\
         \hline
         \hline
2017-06-30  &  \raggedleft 158.00  &  \raggedleft 492  &  89 &  70.76  \\
\hline
2017-12-22  &  \raggedleft 116.70  &  \raggedleft 392  &  80 &  160.92  \\
\hline
2017-06-01  &  \raggedleft 136.88  &  \raggedleft 390  &  70 &  -21.27  \\
\hline
2015-03-31  &  \raggedleft 163.12  &  \raggedleft 387  &  102 &  30.44  \\
\hline
2017-05-31  &  \raggedleft 142.81  &  \raggedleft 387  &  64 &  40.17  \\
\hline
2016-01-29  &  \raggedleft 150.80  &  \raggedleft 386  &  88 &  997.39  \\
\hline
2017-12-01  &  \raggedleft 123.99  &  \raggedleft 378  &  77 &  212.67  \\
\hline
2016-07-15  &  \raggedleft 138.24  &  \raggedleft 371  &  78 &  1,264.70  \\
\hline
2016-06-01  &  \raggedleft 140.73  &  \raggedleft 369  &  92 &  65.04  \\
\hline
2015-01-30  &  \raggedleft 139.19  &  \raggedleft 360  &  55 &  406.47  \\
\hline
2016-12-30  &  \raggedleft 126.59  &  \raggedleft 346  &  97 &  1,104.92  \\
\hline
2017-06-15  &  \raggedleft 130.75  &  \raggedleft 336  &  87 &  57.54  \\
\hline
2017-11-15  &  \raggedleft 108.92  &  \raggedleft 335  &  95 &  56.13  \\
\hline
2015-05-15  &  \raggedleft 128.94  &  \raggedleft 317  &  97 &  24.32  \\
\hline
2017-02-21  &  \raggedleft 120.83  &  \raggedleft 311  &  86  &  206.77  \\
\hline
2015-04-02  &  \raggedleft 111.10  &  \raggedleft 306  &  91 &  52.29  \\
\hline
2015-04-15  &  \raggedleft 133.56  &  \raggedleft 300  &  85 &  40.38  \\
\hline
2015-12-16  &  \raggedleft 153.42  &  \raggedleft 300  &  85 &  14.40  \\
\hline
2015-10-26  &  \raggedleft 123.29  &  \raggedleft 296  &  76 &  299.37  \\
\hline
2017-11-28  &  \raggedleft 91.92  &  \raggedleft 290  &  83 &  371.20  \\
\hline
2017-09-20  &  \raggedleft 167.21  &  \raggedleft 284  &  51 &  33.35  \\
\hline
2017-11-16  &  \raggedleft 107.90  &  \raggedleft 282  &  62 &  56.09  \\
\hline
2017-09-26  &  \raggedleft 101.61  &  \raggedleft 276  &  72 &  213.15  \\
\hline
2017-10-12  &  \raggedleft 92.62  &  \raggedleft 275  &  84 &  97.37  \\
\hline
2016-01-08  &  \raggedleft 103.40  &  \raggedleft 271  &  65 &  170.64  \\
\hline
2015-04-20  &  \raggedleft 117.34  &  \raggedleft 265  &  73 &  241.88  \\
\hline
2015-03-26  &  \raggedleft 111.17  &  \raggedleft 262  &  87 &  167.41  \\
\hline
2015-03-23  &  \raggedleft 112.44  &  \raggedleft 257  &  102 &  60.53  \\
\hline
2015-03-04  &  \raggedleft 117.99  &  \raggedleft 256  &  95 &  160.16  \\
\hline
2015-02-23  &  \raggedleft 101.36  &  \raggedleft 240  &  84 &  542.75  \\
\hline
\hline
\textbf{Average}  &  \raggedleft\textbf{125.76}  &  \raggedleft\textbf{323}  &   \raggedleft\textbf{81}  &  \textbf{239.93}  \\
\hline
\end{tabular}
\label{table:b70_stats_main}
\end{table}


\begin{table}
\centering
\caption{Comparison of the results across the batches of $n=70$ payments on each day in the sample using the order from CQM solver against original FIFO order. Average and median savings are calculated using only the batch of payments that saw improvement (for example, using 89 improved batches on 2017-06-30 in~\autoref{table:b70_stats_main}). The number of worsened batches using CQM settlement compared to FIFO are due to stochastic noise in the quantum solver. In these 30 days, out of the 9,717 batches tested, only 7 come back worse than FIFO (0.072\% error rate). These errors could be easily identified and corrected by re-running the batches.}
    \begin{tabular}{| c |  >{\raggedleft}m{0.11\textwidth} | >{\raggedleft}m{0.11\textwidth} | >{\raggedleft}m{0.12\textwidth} | >{\centering}m{0.11\textwidth} | >{\centering\arraybackslash}m{0.18\textwidth} |}
    \hline
         Date  &   \centering Average Savings (\$Mn)  &  \centering Median Savings (\$Mn)  &  \centering Maximum Savings (\$Mn)  &  \centering Num. Worsened Batches  &  \centering Additional Liquidity Required (\$Mn) \arraybackslash
         \\
         \hline
         \hline
2017-06-30    &  8.24  &  0.51  &  195.91  &  - -  &  - -  \\
\hline
2017-12-22    &  17.76  &  1.02  &  536.12  &  - -  &  - -  \\
\hline
2017-06-01    &  13.37  &  0.57  &  446.99  &  \raggedleft 1  &  \raggedleft -0.0060\arraybackslash  \\
\hline
2015-03-31    &  8.59  &  1.01  &  129.17  &  - -  &  - -  \\
\hline
2017-05-31    &  11.99  &  0.45  &  205.99  &  - -  &  - -  \\
\hline
2016-01-29    &  19.93  &  0.64  &  568.12  &  - -  &  - -  \\
\hline
2017-12-01    &  5.85  &  0.38  &  150.78  &  - -  &  - -  \\
\hline
2016-07-15    &  20.52  &  0.42  &  1,289.03  &  - -  &  - -  \\
\hline
2016-06-01    &  6.20  &  0.31  &  92.39  &  - -  &  - -  \\
\hline
2015-01-30    &  9.48  &  0.48  &  387.78  &  - -  &  - -  \\
\hline
2016-12-30    &  17.00  &  0.43  &  929.20  &  - -  &  - -  \\
\hline
2017-06-15    &  9.31  &  1.71  &  141.24  &  - -  &  - -  \\
\hline
2017-11-15    &  5.60  &  0.36  &  331.91  &  - -  &  - -  \\
\hline
2015-05-15    &  6.52  &  0.68  &  163.31  &  - -  &  - -  \\
\hline
2017-02-21    &  14.79  &  0.46  &  266.57  &  - -  &  - -  \\
\hline
2015-04-02    &  22.44  &  1.40  &  597.34  &  - -  &  - -  \\
\hline
2015-04-15    &  10.71  &  1.45  &  295.93  &  \raggedleft 1  &  \raggedleft -0.0012\arraybackslash  \\
\hline
2015-12-16    &  15.12  &  0.44  &  318.21  &  - -  &  - -  \\
\hline
2015-10-26    &  11.44  &  0.87  &  234.21  &  \raggedleft 1  &  \raggedleft -0.0084\arraybackslash  \\
\hline
2017-11-28    &  13.61  &  1.43  &  200.57  &  - -  &  - -  \\
\hline
2017-09-20    &  21.71  &  2.18  &  331.22  &  - -  &  - -  \\
\hline
2017-11-16    &  14.78  &  0.91  &  219.75  &  - -  &  - -  \\
\hline
2017-09-26    &  13.14  &  0.50  &  361.89  &  \raggedleft 1   &  \raggedleft -2.96\arraybackslash  \\
\hline
2017-10-12   &  8.22  &  0.32  &  171.14  &  \raggedleft 1   &  \raggedleft -0.0076\arraybackslash  \\
\hline
2016-01-08    &  12.68  &  0.62  &  457.88  &  \raggedleft 1  &  \raggedleft -0.0013\arraybackslash  \\
\hline
2015-04-20    &  23.63  &  0.25  &  334.58  &  - -  &  - -  \\
\hline
2015-03-26    &  13.48  &  1.02  &  190.14  &  - -  &  - -  \\
\hline
2015-03-23    &  15.47  &  1.65  &  241.80  &  - -  &  - -  \\
\hline
2015-03-04    &  8.52  &  0.62  &  139.14  &  - -  &  - -  \\
\hline
2015-02-23    &  17.29  &  0.60  &  253.94  &  \raggedleft 1   &  \raggedleft -0.21\arraybackslash  \\
\hline
\hline
\textbf{Average} &  \textbf{13.25}  &  \textbf{0.79}  &  \textbf{339.41}  &  \raggedleft\textbf{0.16\%}  &  \raggedleft\textbf{-0.11}\arraybackslash  \\
\hline
\end{tabular}
\label{table:b70_stats}
\end{table}

As summarized in~\autoref{table:b70_stats_main}, the quantum solver can provide significant end-of-day liquidity savings---averaged at \$239.9M with a median of 128.8M across 30 days and following an exponential distribution. The most significant savings is \$1.26B on 2016-07-15. However, out of the 9,717 batches examined across these days, 49\% have no participants who moved their mNDP, and thus no rearrangement could provide liquidity savings. 
25.3\% of batches have an optimized rearrangement that provided liquidity savings. The remaining 25.7\% have participants that grew their mNDP but are not optimizable for reasons such as the participants having only outgoing payments, FIFO already being an optimal solution, or a large payment that could not be offset. 

As summarized in~\autoref{table:b70_stats}, on a typical day, an optimized batch saves \$13.25M on average, with a median of \$0.79M and a maximum of \$339.41M. Overall, the maximum savings achieved by any one batch is \$1.29B on July 15, 2016.
Occasionally, the order returned by the CQM hybrid solver results in less efficient liquidity usage than FIFO.  This occurs only seven times out of the 9,717 batches and never occurs more than once on any day.  The loss in savings is less than \$10k in 5 of those instances and reaches as high as \$2.96M (\autoref{table:b70_stats}).  These instances can be attributed to stochastic noise in the quantum annealer and could easily be removed in a practical implementation of this system by immediately re-running the affected batches (if time allowed) or defaulting to FIFO.

\begin{figure}[h!]
\centering
\includegraphics[width=0.7\textwidth]{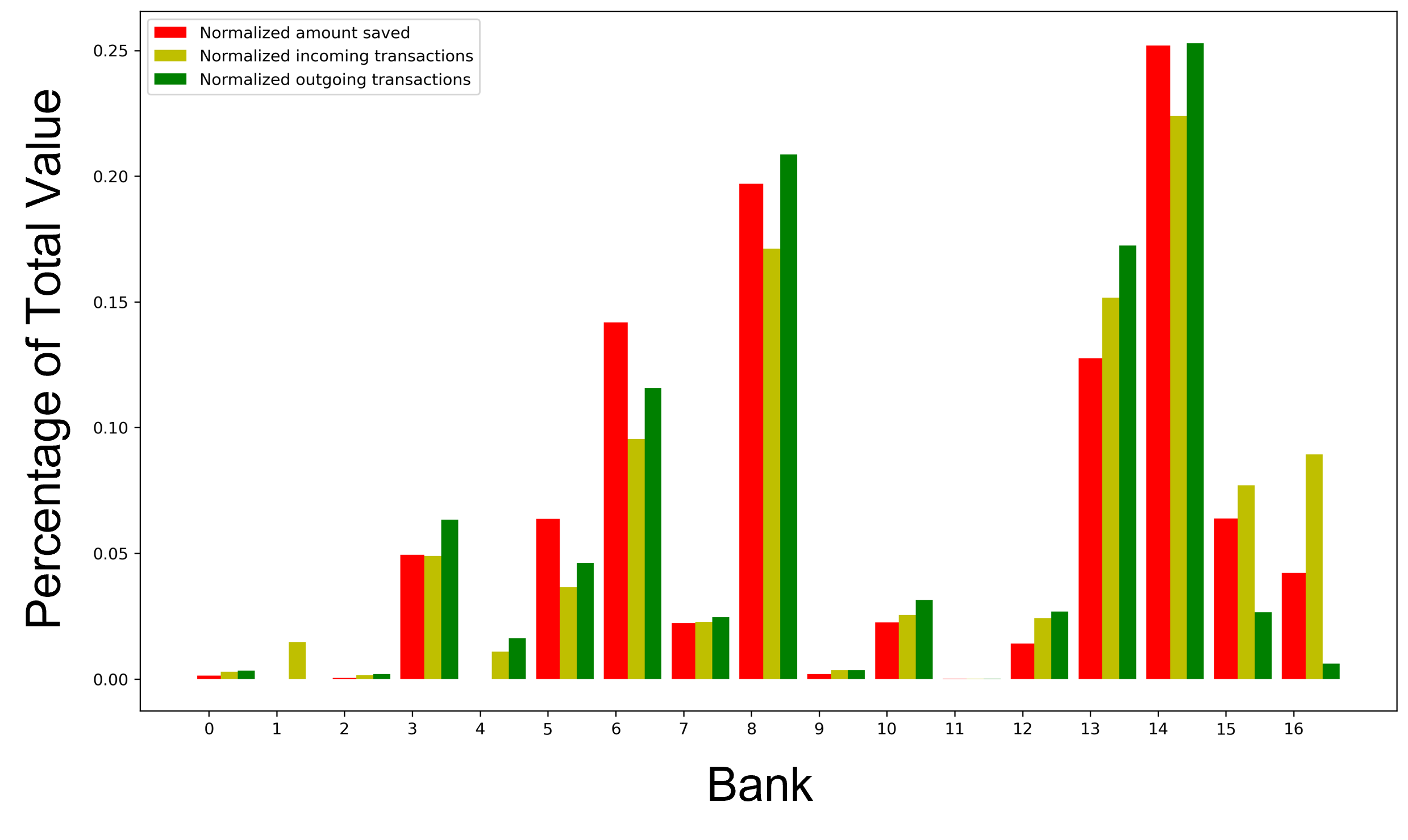}
\caption{Amount of liquidity saved per bank as a percentage of total savings incurred (red) compared with the percentage of the total value of incoming (gold) and outgoing (green) payments using a fixed batch size.}
\label{fig:bank_savings}
\end{figure}

To examine how equally the savings were distributed across participants, in~\autoref{fig:bank_savings} we compare each participant's liquidity savings (using a batch size of 70) to their total transaction participation. The percentage of the total amount saved for each participant is compared to their percentage of total incoming and outgoing transactions. The Pearson's correlation coefficient between percent saved and percent of incoming and outgoing transactions is r=0.961 and 0.967, respectively. This high correlation implies the savings are somewhat fairly distributed based on how much liquidity each participant moves. In other words, if a participant is involved in payments responsible for half of the total settled value in a day, they will likely see about half the savings achieved for that day.

\begin{figure}[!h]
  \centering
    \includegraphics[width=0.493\textwidth]{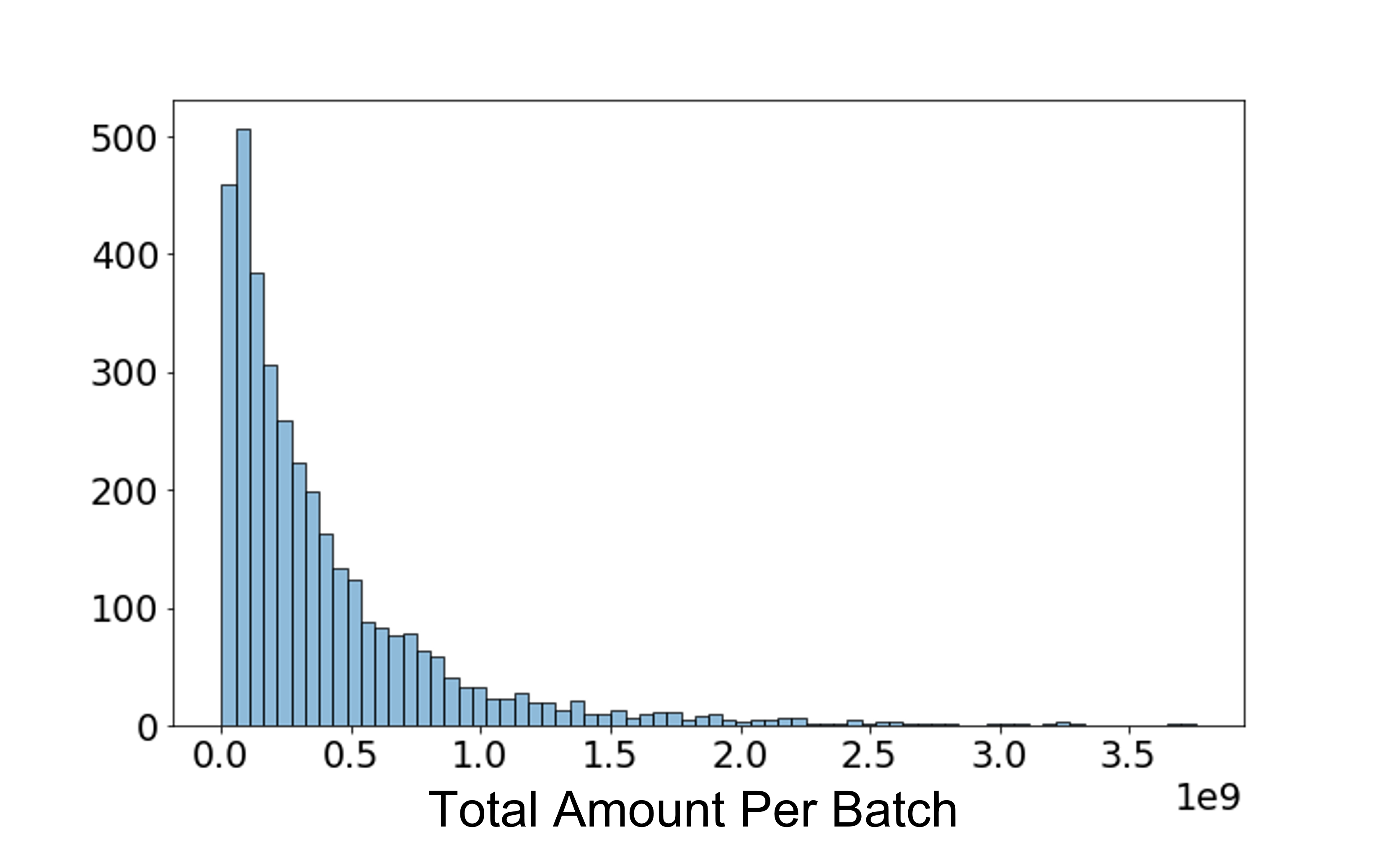}
    \includegraphics[width=0.493\textwidth]{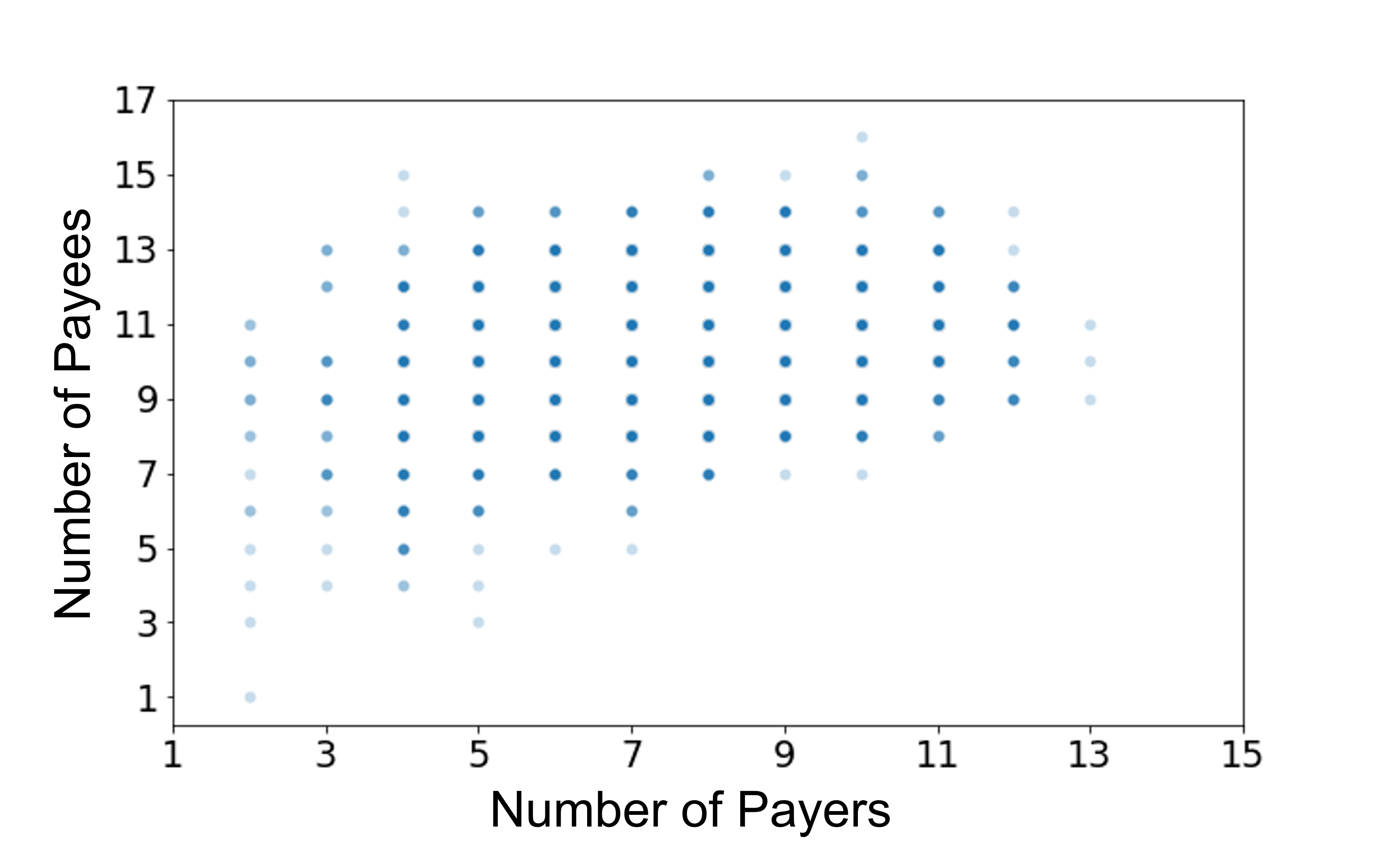}
    \caption{(Left) total value (in CAD) settled per batch in our sample. (Right) the scatter plot of the number of unique payers and payees involved in each batch in our sample (darker blue dots show more repetitions and lighter blue dots shows fewer repetitions).}
  \label{fig:value_payers_per_batch}
\end{figure}

In most cases, including ours, the payment values follow a Pareto distribution, maintained in the total settled values for each batch (\autoref{fig:value_payers_per_batch}, left). In our sample, the average value settled for a batch size of 70 is \$0.4 billion, but some settle for \$3.5 billion. The non-normal distribution of payments makes it difficult to find bilateral or multilateral netting opportunities. To further reduce the optimization capability, most batches have more payees than payers. \autoref{fig:value_payers_per_batch} (right) shows the number of payees and payers per batch using transparency to represent their frequency. On average, there are eleven payers and eight payees. The presence of a lone payee in batches acts as a liquidity sink removing it from the liquidity pool. We infer that a strongly interconnected network of equal value payments in each batch would provide an opportunity for larger liquidity savings. 

\subsection{Performance of Quantum Algorithm on a Chosen Day}
\begin{figure}[h]
\centering
\includegraphics[width=0.65\textwidth]{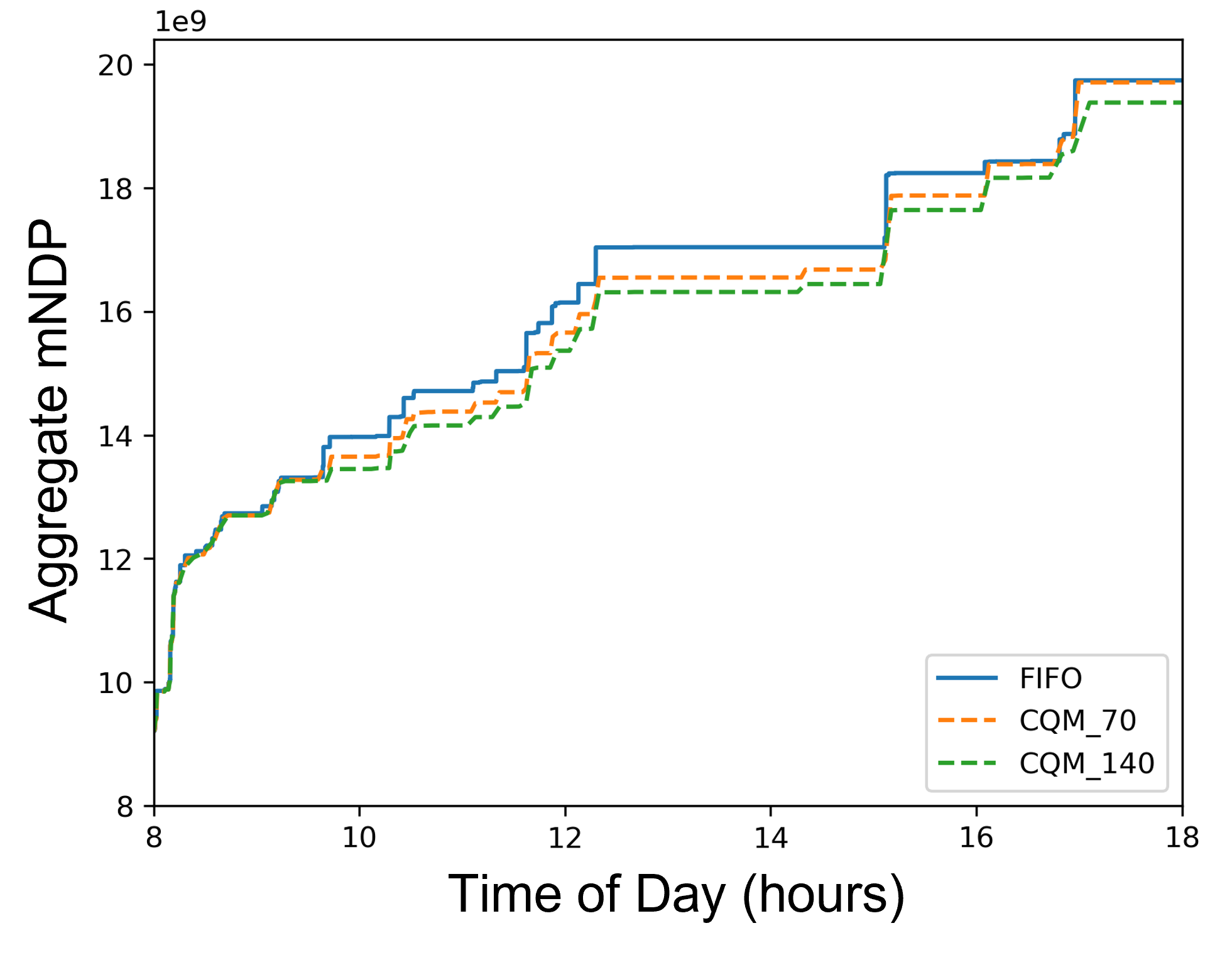}
\caption{The aggregate max net debit position (mNDP) versus time of the day for Sept 20, 2017. The blue line shows the original payment order (FIFO); the orange and green lines respectively show the orders returned from the quantum annealers,  with the day divided into batch sizes of 70 and 140 payments.}
\label{fig:2017-09-20_AmNDP}
\end{figure}

Next, as an example, we show the evolution of the aggregate mNDP for a typical day (Sept 20, 2017) in our sample in~\autoref{fig:2017-09-20_AmNDP} for a batch size of $n=70$. At the start of the day, as participants send their payments, we see a sharp rise in aggregated mNDP. As the day progresses, however, we see a slower increase in aggregated mNDP. For many batches during the day, no participant had moved their mNDP, and thus no rearrangement could provide liquidity savings. These are seen as flat sections in the aggregate mNDP curve in ~\autoref{fig:2017-09-20_AmNDP}.

Sometimes, a large payment (or group of payments) in a single batch can cause the aggregate mNDP to converge for the CQM solver and FIFO, erasing savings that the CQM had achieved earlier in the day. Examples of this can be seen at 14:20 and more extremely at 16:05 in \autoref{fig:2017-09-20_AmNDP}. This erasure cannot be avoided without increasing the queue size or using a more complicated queue-building algorithm that allows payments to move between queues to create larger graph cycles and offset the required liquidity needed to settle the large payment(s). Clearly, for a chosen batch size, there is significant heterogeneity across the batches, and there are limitations on the improvements which can be achieved with this simple pre-processing setup.

\begin{figure}[!h]
  \centering
    \includegraphics[width=0.9\textwidth]{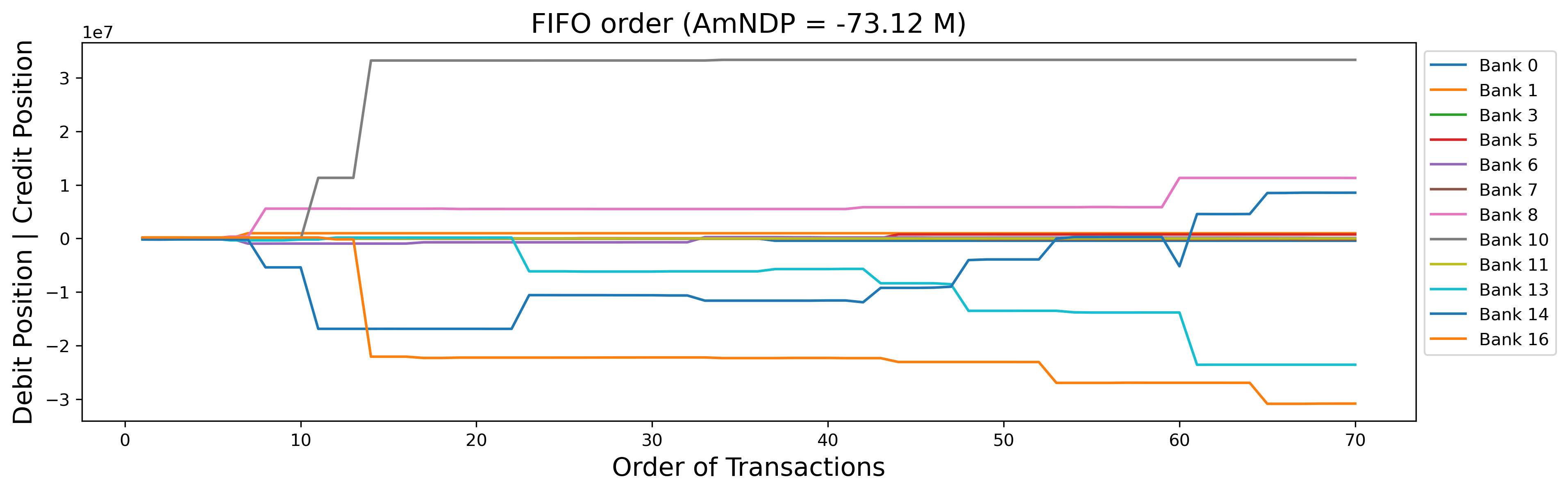}
	\includegraphics[width=0.9\textwidth]{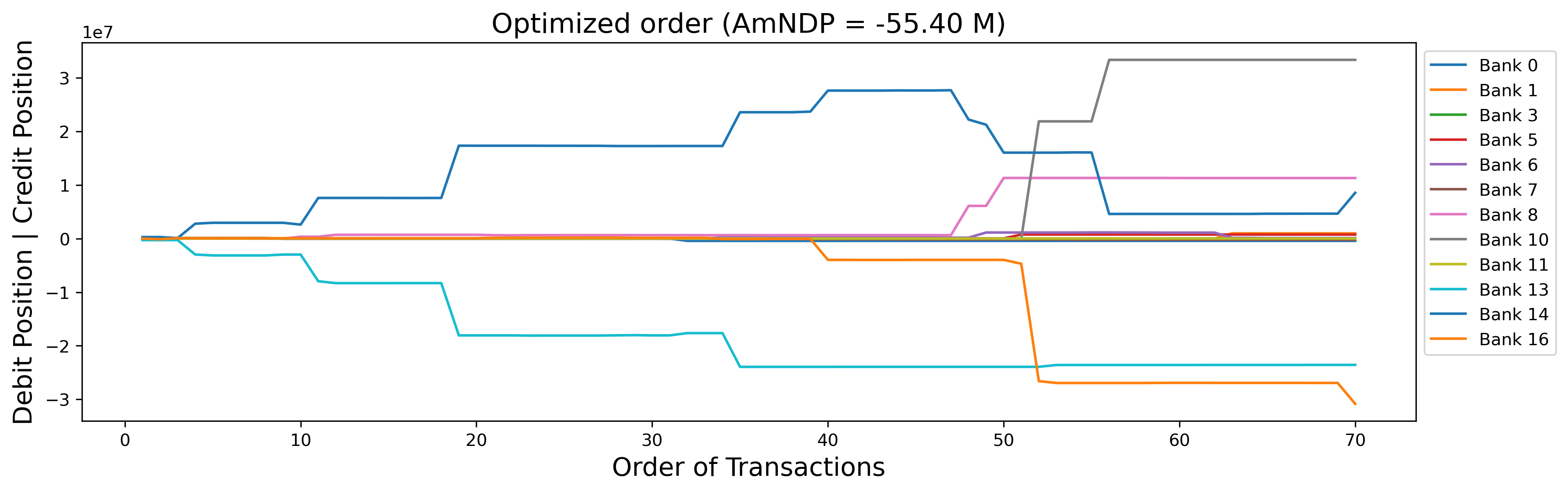}
    \caption{The change in individual banks' credit or debit position for a given queue of 70 transactions immediately after each payment settlement for Sept 20, 2017. The top is for FIFO order (\$73 million aggregate mNDP), and the bottom is for the order proposed by the hybrid quantum CQM solver (\$55 million aggregate mNDP).}
  \label{fig:order_balance}
\end{figure}

To help visualize the queue reordering process in a typical batch of 70 payments (on Sept 20, 2017), each participant's debit and credit balance evolution is shown in \autoref{fig:order_balance}. The top chart uses the FIFO order, and the bottom chart is for the order proposed by quantum optimization. Each line represents each bank's balance as the transactions are completed sequentially in the prescribed order. Since we optimize the aggregate mNDP, we want to limit the depth to which the participants drop below zero. In this example, the mNDP for Bank 14 was originally \$17 million. The optimization rearranges the transactions to prevent Bank 14 (blue line) from taking any debit position while not penalizing other participants. Doing so, and reordering other banks' payments, saves the system 18 million in aggregate mNDP over these 70 transactions. However, it should be noted that not all transactions in this batch can be optimized. Two examples are banks 13 (teal) and 16 (orange), which have only outgoing payments; therefore, no scope exists for optimizing their mNDP through queue reordering any of their transactions in this batch.

\begin{figure}[h!]
\centering
\includegraphics[width=0.9\textwidth]{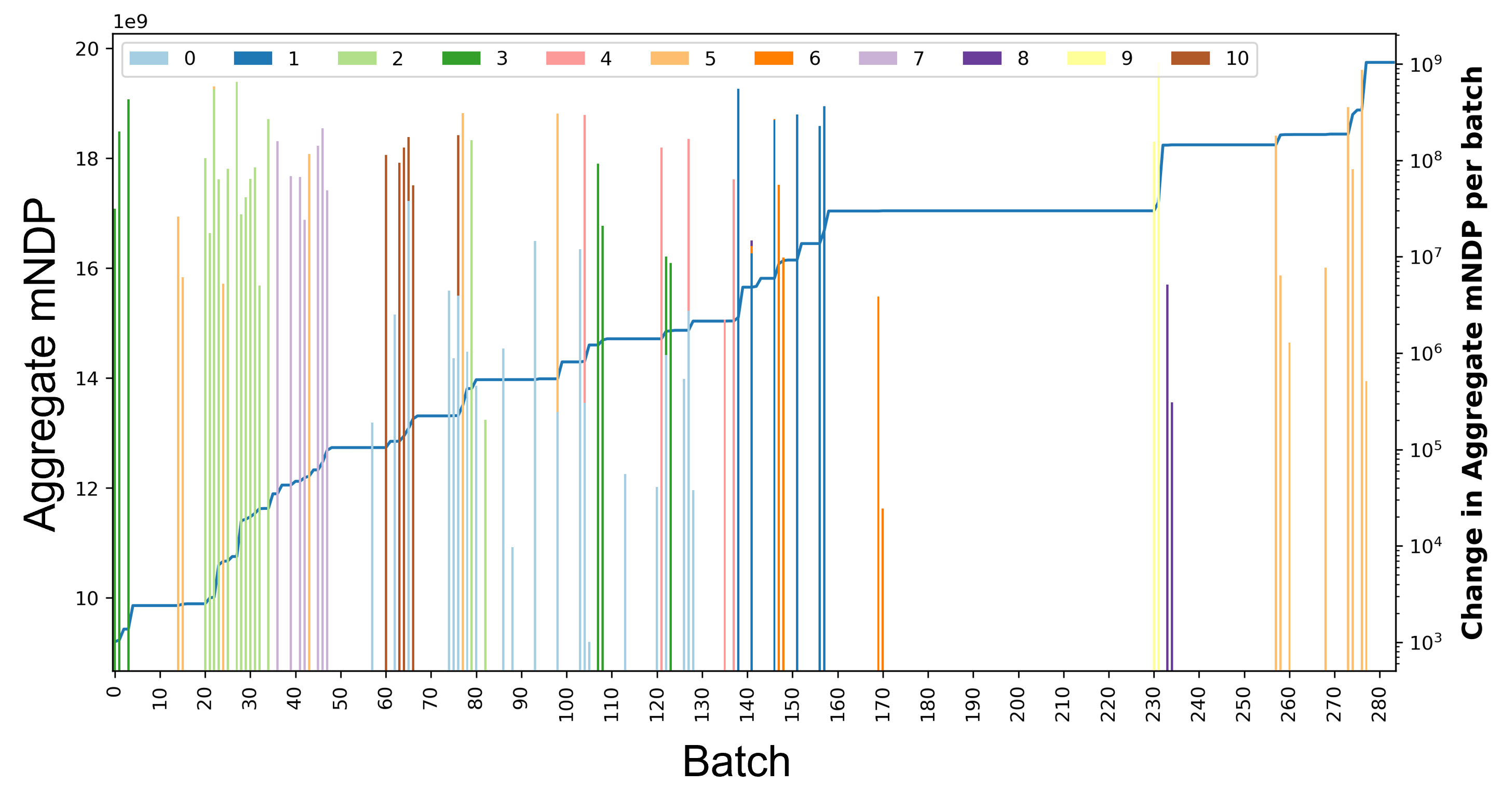}
\caption{The change in aggregate max net debit position against the accumulated batches of payments for Sept 20, 2017.  The blue line represents the aggregated mNDP (on the left axis) and stacked bars represents magnitude of the change in mNDP caused by the individual participants (on the right axis).}
\label{fig:2017-09-20_AmNDP_change}
\end{figure}

To understand the network effects in terms of the number of payees and payers per batch, in~\autoref{fig:2017-09-20_AmNDP_change} we show the participants that cause the aggregate mNDP to increase (if at all) for each batch as a stacked bar plot with FIFO settlement on Sept 20, 2017. The increase is usually the result of only one participant (a notable exception is batch 98, for example). Therefore, minimizing these participants' mNDP would provide the most significant liquidity savings. However, the queue rearrangement can do little if the participants have a large negative change to their net position in a single batch, i.e., they do not have incoming payments to offset their one or more large outgoing payments. For this reason, FIFO and CQM with 70 and 140 queue size track each other with varying offsets and can exhibit the convergence that erases gains discussed above (see~\autoref{fig:2017-09-20_AmNDP}). Also evident are groupings throughout the day where one participant moves the aggregate mNDP in succession. These are cycles where participants have repeated outgoing transactions and a low balance. 

Finally, we run the algorithm by doubling the batch size to $n=140$ for two test days: Sept 20 and Oct 12, 2017.
Similar to~\autoref{table:b70_stats_main}, we run a horse race between the FIFO and CQM, where they both start with the same initial conditions and settle every batch throughout the day using FIFO and CQM order, and at the end of the day, we report the mNDP.
As detailed in \autoref{table:b140}, the increase in batch size produces substantial improvements in savings, as expected.  Most notably, the number of batches that see higher savings over FIFO increases by close to 15\% for both days with respect to the number of batches that could potentially be optimized.  This is certainly due to the increase in both the amount of liquidity available for recycling and the increased connectivity between payers and payees in a single batch's network graph. 


\begin{table}[h]
\centering
    \caption{Comparison of the results for two days using the order proposed by the CQM hybrid solver and the original order (FIFO) with batch size $n=140$. The improvements over the same method using the the batch size of $n=70$ are shown in bold.}
\resizebox{\textwidth}{!}{
    \begin{tabular}{| c | >{\raggedleft}m{0.15\textwidth}| >{\raggedleft}m{0.15\textwidth} | >{\raggedleft}m{0.15\textwidth} | >{\raggedleft}m{0.15\textwidth} | >{\raggedleft}m{0.15\textwidth} | >{\raggedleft\arraybackslash}m{0.15\textwidth} |  >{\raggedleft\arraybackslash}m{0.15\textwidth} |}
        \hline
        Date  &  \centering Num. of Batches  &   \centering Optimizable Batches  &  \centering  Num. Improved Batches  &  \centering Average Savings (\$1M)  &  \centering Median Savings (\$1M)  &  \centering Maximum Savings (\$1M)  &  \centering End-of-Day Savings (\$1M)\arraybackslash  \\
        \hline
2017-09-20  &  142  &  87  &  50   &  31.46  &  5.93  &  508.41  &  359.56  \\
\hline
\multicolumn{3}{|c|}{\textbf{Improvement Over n=70:}}  &  \textbf{+13.9\%}  &  \textbf{+9.75}  &  \textbf{+3.75}  &  \textbf{+177.19}  &  \textbf{+326.21}  \\
\hline
2017-10-12  &  138  &  110  &  73   &  16.84  &  1.16  &  171.15  &  191.84  \\
\hline
\multicolumn{3}{|c|}{\textbf{Improvement Over n=70:}}  &  \textbf{+14.5\%}  &  \textbf{+8.62}  &  \textbf{+0.84}  &  \textbf{+0.01}  &  \textbf{+94.47}  \\
\hline
    \end{tabular}
    }
    \label{table:b140}
\end{table}

On Sept 20, 2017, we see an impressive improvement in average, median, and maximum savings of approximately 45\%, 172\%, and 53\%, respectively. On Oct 12, 2017, we see an even greater improvement in its average and median savings of approximately 105\% and 263\%, respectively, while its maximum savings see a negligible improvement of less than 1\%.  However, maximum savings as a performance metric is the most dependent on how (and if) the day's largest payments are grouped into the queues, and it is expected to be the least consistent of the three measures.

Arguably, the most encouraging improvement is in the end-of-day savings, where these two days see an increase of approximately \$326M (978\%) and \$192M (97\%), respectively.  These dramatic changes can be attributed to both the overall superior performance of the larger queue size, as well as a reduction in the loss of mid-day savings.  This is particularly evident for 20 Sept, 2017, also depicted in~\autoref{fig:2017-09-20_AmNDP}.

\section{Conclusions and Future Work}\label{sec:conclusions}


We show that substantial liquidity savings can be achieved with the addition of a pre-processing layer where payments are placed in a queue and reordered using an algorithm that makes use of hybrid quantum-classical resources before being submitted to the RTGS for settlement.  Using conventional computing hardware, optimizing the queue requires prohibitively long computation times to achieve results for all but trivial queue sizes. However, quantum computing via hybrid annealing reduces computation times such that queues that are large enough to provide significant savings potential can be optimized. In a sample of 30 representative days of transactions from Canada's HVPS, we demonstrate such improvements for queue sizes of 70 and 140, with average aggregate daily liquidity savings of C\$239.93M and C\$275.70M, respectively. Furthermore, these savings tend to be distributed proportionately to the participants' payments values.

Although much larger savings are achieved on an intraday basis (one batch reached nearly C\$1.3B), the limit in the potential savings is set by the timing and structure of payment flows between participants, especially when large payments appear in queues with few or no other payments involving that payer. In such cases, only minimal improvements are possible. Such challenges suggest avenues for future improvements,  for example, combining a bypass component (used today in traditional LSMs) that sets aside larger payments, allowing other payments to be optimized. Likewise, the progress of quantum technology will permit optimizing larger queue sizes in the future, which could help overcome such limitations. Lastly, we may extend our algorithm to implement a flexible batch size, based on a combination of queue time, the value to process, and the balance of payers and payees, to improve the chances of offsetting payments occurring in the same batch.  


In the end, we found the answer that we sought:  quantum technologies readily available today \emph{can} be used to optimize the order of payments.  While the trade-off of liquidity and delay remains, the small delay required by our quantum pre-processor can provide substantial liquidity savings in payment systems.  Ultimately, we demonstrate that a quantum method could reduce liquidity usage without significantly increasing settlement delay, and this paves the way for future research as these technologies mature. 

\subsection*{PIVOT}
\noindent
Through the Partnerships in Innovation and Technology Program (PIVOT), the Bank of Canada works with innovators in the private sector and academia to experiment with digital tools and technologies.  This report is the result of a PIVOT collaboration between the Bank and GoodLabs Studio, a startup with expertise in fintech, AI, HPC, and other emerging technologies.  

\clearpage
\bibliographystyle{unsrt}  
\bibliography{references}


\appendix

\section{Formulation of the Objective Function}
\label{app:H}

Queue optimization is done analytically through minimizing an objective function. A number of payments,  $n$, are collected in a queue. The payments are ordered with index $i=1\ldots n$, while the rearranged queue has the alternative index $t=1\ldots n$. The optimization of the Hamiltonian returns the mapping $i\rightarrow t$ that minimizes the amount of liquidity needed in the system (the aggregate max net debit position). We define a participant $\alpha$'s net position in the final queue at position $t$  as $N_\alpha(t)$.  Each participant's maximum net debit position incurred from previous batches is included as $mNDP_\alpha$ (zero for the initial batch). Thus, the collateral needed by each participant to ensure that they do not go into a negative liquidity position is

\begin{equation}
    {mNDP}_\alpha=|\min\{N_\alpha(\text{previous\ queues}),0\}|.
\end{equation}

The participant can access liquidity from either $N(t)$ (if their position happens to be a net credit) or this collateral. More precisely, their total liquidity available to settle payment $t$ is effectively
\begin{equation}
    \ell_\alpha(t)=N_\alpha(t)+ {mNDP}_\alpha.
\end{equation}

A participant's instantaneous net position can be calculated using
\begin{equation}
N_\alpha(t) = N_\alpha(0) + \sum_{i=1}^{n}\sum_{\tau=1}^{t} f(\alpha,i)\,x_{i,\tau},
\end{equation}
where $N_\alpha(t)\geq -{mNDP}_\alpha$, $N_\alpha(0)$ is participant $\alpha$'s net position immediately before the settlement of this batch (starting balance), $x_{i,\tau}$ is a binary decision variable that indicates whether payment $i$ is settled ($x_{i,\tau}=1$) or not ($x_{i,\tau}=0$) at position $\tau$ in the final queue,
\begin{equation}
f(\alpha,i) =
\begin{cases}
\ v,&\text{if }\alpha\text{ is the payee of payment }i\\
-v,&\text{if }\alpha\text{ is the payer of payment }i\\
\ 0,&\text{otherwise}
\end{cases},
\end{equation}
and $v$ is the value of payment $i$.

Ideally, we seek a mapping $i\rightarrow t$ such that $\ell_\alpha(t)\geq 0\ \forall \alpha,t$.  However, such a mapping may not exist for a given batch.  For the general case, we define the positive real number $b_\alpha$ that is a hypothetical amount of liquidity one would need to add to participant $\alpha$'s account in order to keep their balance non-negative for all payments $t$.  We now have the inequality
\begin{equation}
b_\alpha + N_\alpha(0) + {mNDP}_\alpha + \sum_{i=1}^{n}\sum_{\tau=1}^{t} f(\alpha,i)\,x_{i,\tau}\geq 0
\label{eq:bal_constr}
\end{equation}
that can always be satisfied for the trivial case $b_\alpha\rightarrow\infty$.  The optimal mapping occurs when $b_\alpha$ is collectively minimized across $\alpha$.  Quite simply,
\begin{equation}
\min_{b_\alpha,x_{i,t}}\sum_\alpha b_\alpha.
\end{equation}
Note that the optimal $b_\alpha$ represents the required change that will have to be made to $mNDP_\alpha$ before the next queue is optimized.  Equivalently, this is the additional collateral that the central bank will demand from the participant $\alpha$ to provide the liquidity before it allows settlement of the current queue to commence.

This problem is of the mixed binary optimization (MBO) type where the optimization is over both continuous ($b_\alpha$) and binary ($x_{i,t}$) variables.  We can easily transform this into a quadratic unconstrained binary optimization (QUBO) problem for processing on a quantum device~\cite{Glover2019,Lucas2014}. First, the inequality constraint given in ~\autoref{eq:bal_constr} must be written as an equality constraint with the addition of a slack variable $s_\alpha(t)$:
\begin{equation}
b_\alpha+N_\alpha(0)+\text{mNDP}_\alpha+\sum_{\tau}^{t}\sum_{i=1}^{\tau} f(\alpha,i)\,x_{i,\tau}-s_\alpha(t)=0
\end{equation}
where $s_\alpha(t)\geq 0$ is an additional continuous variable over which optimization must occur.  Second, this constraint must be enforced in the QUBO as a quadratic penalty term with a Lagrange multiplier.  Third, all continuous variables ($b_\alpha$, $s_\alpha(t)$) must be discretized to the nearest cent (or larger base value with the loss of some resolution) and then represented with the so-called ``log trick.''  For example, $b_\alpha$ is replaced with the sum of a geometric series of binary variables $\beta_\alpha$,
\begin{equation}
b_\alpha = \sum_j 2^j\beta_\alpha^{(j)},
\end{equation}
where $\beta_\alpha^{(0)}$ represents the least-significant digit in the binary representation of $b_\alpha$, $\beta_\alpha^{(1)}$ the second-least-significant digit, etc.

Finally, two more constraints must be added to ensure that every payment is settled exactly once; to guarantee the mapping $i\rightarrow t$ is one-to-one.  To ensure that every $i$ corresponds to exactly one $t$, we have
\begin{equation}
    1-\sum_t x_{i,t} = 0\ \forall i,
\end{equation}
and similarly that every $t$ corresponds to exactly one $i$, we have
\begin{equation}
    1-\sum_i x_{i,t} = 0\ \forall t.
\end{equation}

In the end, we are left with the equation
\begin{align}
H=& \sum_\alpha b_\alpha + \sum_{\alpha,t} \lambda_1 \left(b_\alpha + N_\alpha(0) + \text{mNDP}_\alpha + \sum_i\sum_{\tau\leq t} f(\alpha,i)\, x_{i,\tau} - s_\alpha(t) \right)^2 \nonumber \\
& + \sum_i \lambda_2 \left( 1-\sum_t x_{i,t}\right)^2 + \sum_t \lambda_2\left(1-\sum_i x_{i,t} \right)^2,
\label{eq:H}
\end{align}
where the log trick has been omitted for brevity.  Minimizing this Hamiltonian over all variables
\begin{equation}
    \min_{b_\alpha,\,x_{i,t},\,s_\alpha(t)} H
\end{equation}
yields the optimal reordering of the payment queue.  (Note that the same Lagrange multiplier, $\lambda_2$, can be assumed for both one-to-one constraints without loss of generality due to their symmetry.)

\end{document}